\documentclass[sort&compress]{elsarticle}
\usepackage{amsmath}
\usepackage{graphicx}
\usepackage{color}
\usepackage{natbib}
\usepackage{hyperref}

\newcommand{\dd}{\mathrm{d}}

\begin{document}
% \begin{flushleft}
% TTP13-0XX\\
% SFB/CPP-13-XX\\
% arXiv:YYMM.NNNN
% \end{flushleft}
\begin{frontmatter}
\title{The QED vacuum polarization function at four loops and the
  anomalous magnetic moment at five loops} 
\author[Moskau]{P.A.~Baikov}
\author[Muenchen]{A.~Maier}
\author[Zeuthen]{P.~Marquard} \address[Moskau]{Skobeltsyn Institute of Nuclear Physics, Lomonosov Moscow State University,
1(2), Leninskie gory, Moscow 119991, Russian Federation}
\address[Muenchen]{Physik
Department T31, James-Franck-Stra\ss{}e, Technische Universit\"at M\"unchen, D–85748 Garching, Germany} 
% \address[Karlsruhe]{Institut f\"ur Theoretische
%   Teilchenphysik, Karlsruhe~Institute~of~Technology~(KIT), 76128
%   Karlsruhe, Germany}
\address[Zeuthen]{Deutsches Elektronen Synchrotron DESY, Platanenallee 6, D15738 Zeuthen, Germany}
% \begin{flushleft}
% TTP13-0XX\\
% SFB/CPP-13-XX\\
% arXiv:YYMM.NNNN
% \end{flushleft}
\begin{abstract}
  The anomalous magnetic moment of the muon is one of the most fundamental
  observables. It has been measured experimentally with a very high
  precision and on theory side the contributions from perturbative QED
  have been calculated up to five-loop level by numerical
  methods. Contributions to the muon anomalous magnetic moment from
  certain diagram classes are also accessible by alternative methods. In
  this paper we present the evaluation of contributions to the QED
  corrections due to insertions of the vacuum polarization function at
  five-loop level.
\end{abstract}
\end{frontmatter}

\section{Introduction}
\label{sec:intro}
The anomalous magnetic moments of electron and muon are among the best
experimentally measured quantities and are also very well understood from
theoretical side. In particular, the perturbative QED corrections have
been under consideration for a very long time. The leading-order QED
corrections have been considered by Schwinger in
Ref.~\cite{Schwinger:1948iu}, the next-to-leading order corrections not
much later in Refs.~\cite{Petermann:1957hs,Sommerfield:1957zz}. At three
loops, i.e. next-to-next-to-leading order, the QED corrections have been
first calculated numerically in Ref.~\cite{Kinoshita:1995ym} and later
analytically in
Refs.~\cite{Laporta:1996mq,Melnikov:2000qh,Marquard:2007uj}.  At four and
five loops the complete QED contributions have been calculated
numerically in Refs.~\cite{Aoyama:2012wk,Aoyama:2012wj} while only
partial analytical results
exist~\cite{Lautrup:1974ic,Kinoshita:1990ur,Kawai:1991wd,Faustov:1990zs,Kataev:1991cp,Broadhurst:1992za,Baikov:1995ui,Laporta:1993ds,Baikov:2012rr,Lee:2013sx}. For
a more thorough review of the current status see
e.g. Ref.~\cite{Jegerlehner:2009ry}.

In the case of the muon anomalous moment, next to the universal
contribution, contributions from diagrams with closed electron loops are
of particular interest. A subgroup of this class of diagrams,
contributions from corrections to the photon propagator, can be easily
obtained using already available building blocks
\cite{Lautrup:1974ic,Barbieri:1974nc}. At five loops the leading
contributions have been obtained from considering the asymptotic form~\cite{Baikov:2012rr}, which resulted in unexpected discrepancies with
the results obtained in Ref.\cite{Aoyama:2010zp}. The aim of this paper
is to improve the predictions for the anomalous magnetic moment of the
muon made in Ref.~\cite{Baikov:2012rr} and to resolve the
discrepancies. To this extend we approximately reconstruct the photon
vacuum polarization at four loops using all available information.

This paper is organized as follows: In
Section~\ref{sec:qed-vacu-polar} we collect the necessary information
to construct an approximation for the vacuum polarization function at
four loops. In Section~\ref{sec:muon} we present the calculation of the
contribution to the anomalous magnetic muon from vacuum polarization
insertions. 

\section{QED vacuum polarization at four loops}
\label{sec:qed-vacu-polar}
We define the vacuum polarization function $\Pi(q^2)$ as usual by
\begin{equation}
  \label{eq:4}
  (q^\mu q^\nu - q^2 g^{\mu\nu}) \Pi(q^2) = i e^2 \int \dd x \,\langle 0 | e^{i q x} T j^\mu(x) j^\nu(0) | 0 \rangle   \,,
\end{equation}
with the current $j^\mu = \bar\psi \gamma^\mu \psi$ and write it as an
expansion in the fine-structure constant $\alpha$
\begin{equation}
  \label{eq:5}
  \Pi(q^2) = \frac{\alpha}{\pi}\Pi^{(1)}(q^2) 
+\left(\frac{\alpha}{\pi}\right )^2\Pi^{(2)}(q^2) 
+\left(\frac{\alpha}{\pi}\right )^3\Pi^{(3)}(q^2) 
+\left(\frac{\alpha}{\pi}\right )^4\Pi^{(4)}(q^2) 
+{\cal O}(\alpha^5)  \,.
\end{equation}
At three and four loops the vacuum polarization can only be calculated
in a low-energy, high-energy or threshold expansion at the moment.  The
available information can then be used to construct an approximate
function using Pad\'e approximation. In the following we will present
the results for the corresponding expansions and construct an
approximating function. The renormalization of lepton mass and
fine-structure constant $\alpha$ is performed in the on-shell
scheme. Note that in this scheme the condition $\Pi(0) = 0$ is being
imposed.

\subsection{Low-energy expansion}
\label{sec:low-energy-expansion}
In the low-energy limit the polarization function can be expanded in a
power series in $z=q^2 / ( 4m_q^2 ) < 1$
\begin{equation}
  \label{eq:6}
  \Pi^{(n)}_{\mathrm{le}} = \sum_{k=1} \Pi^{(n)}_{\mathrm{le},k} z^k \,.
\end{equation}
For the QED case at hand, the result for the non-singlet
contribution can in principle be obtained from the QCD results given in
Ref.~\cite{Maier:2009fz}. But since the results are not given expressed
through colour factors they cannot easily be translated to the case of
QED. The low-energy expansion was therefore recalculated for the case at
hand. The calculation follows a well-established path. The Feynman
diagrams are generated using {\tt qgraf}~\cite{Nogueira:1993} and mapped
onto six topologies using {\tt q2e} and {\tt
exp}~\cite{Seidensticker:1999bb,Harlander:1997zb}. Then a {\tt
FORM}~\cite{vermaseren-form} program is used to apply projectors and
take traces. The resulting scalar integrals are then reduced to master
integrals using {\tt Crusher}~\cite{crusher}, which implements Laporta's
algorithm~\cite{Laporta:2001dd} for solving integration-by-parts
identities. The needed master integrals have been calculated in
Refs.~\cite{Chetyrkin:2006dh,Schroder:2005va,Schroder:2005hy,Laporta:2002pg,Chetyrkin:2004fq,Kniehl:2005yc,Schroder:2005db,Bejdakic:2006vg,Kniehl:2006bf,Kniehl:2006bg}. Combining
all steps and performing the renormalization of all quantities in the
on-shell scheme, the first three moments are obtained { \small
  \begin{equation}
    \label{eq:234}
    \begin{split}
&\Pi^{(4)}_{\mathrm{le},1} =
\Bigg(-\frac{30853 a_1^4}{22680}+\frac{30853 \pi ^2 
a_1^2}{22680}-\frac{64 \pi ^2 a_1}{135}-\frac{30853 
a_4}{945}-\frac{77255063 \zeta_3}{4762800}  \\
&+\frac{473237 \pi 
^4}{1360800}+\frac{866 \pi ^2}{1215}-\frac{3619201}{529200}\Bigg) 
n_l^2+\Bigg(\frac{3376 a_1^5}{2835}-\frac{2178299 
a_1^4}{544320}-\frac{3376 \pi ^2 a_1^3}{1701} \\
&+\frac{2178299 \pi ^2 
a_1^2}{544320}-\frac{12191 \pi ^4 a_1}{8505}-\frac{1922 \pi ^2 
a_1}{405}-\frac{2178299 a_4}{22680}-\frac{27008 a_5}{189}+\frac{\pi ^2 \zeta
_3}{30}\\
&-\frac{69694097 \zeta_3}{1209600}+\frac{33931 \zeta_5}{189}+\frac{66975707 \pi ^4}{65318400}+\frac{7297 \pi 
^2}{3240}+\frac{3562169}{2721600}\Bigg) n_l+ \\
&\mathbf{si} \Bigg(-\frac{739 
a_1^4}{1080}+\frac{739 \pi ^2 a_1^2}{1080}-\frac{739 
a_4}{45}-\frac{2017831 \zeta_3}{213840}+\frac{35 \pi 
^4}{216}+\frac{664837}{641520}\Bigg) n_l^2 \\
&+\Bigg(-\frac{1291 \zeta
_3}{1944}-\frac{32 \pi ^2}{2025}+\frac{83971}{78732}\Bigg) n_l^3 \,,
\end{split}
\end{equation}
\begin{equation}
\begin{split}
&\Pi^{(4)}_{\mathrm{le},2} =
\Bigg(-\frac{130829911 a_1^4}{8709120}+\frac{130829911 \pi ^2 
a_1^2}{8709120}-\frac{128 \pi ^2 a_1}{315}-\frac{130829911 
a_4}{362880}\\
&-\frac{5763324918049 \zeta_3}{26824089600}+\frac{2208846791 \pi 
^4}{522547200}+\frac{10846 \pi 
^2}{14175}-\frac{42466787908001}{1086375628800}\Bigg) 
n_l^2 \\
&+\Bigg(\frac{19593724 a_1^5}{2027025}-\frac{60530131639 
a_1^4}{24908083200}-\frac{19593724 \pi ^2 
a_1^3}{1216215}+\frac{60530131639 \pi ^2 
a_1^2}{24908083200}\\
&-\frac{72540947 \pi ^4 a_1}{6081075}-\frac{3058 \pi ^2 
a_1}{675}-\frac{60530131639 a_4}{1037836800}-\frac{156749792 
a_5}{135135}+\frac{\pi ^2 \zeta_3}{35}\\
&-\frac{112244692092317 \zeta
_3}{6974263296000}+\frac{14315837 \zeta_5}{10010}-\frac{427149037853 \pi 
^4}{2988969984000}+\frac{839 \pi 
^2}{378}+\frac{94330906317547}{3487131648000}\Bigg) n_l\\
&+\mathbf{si} 
\Bigg(\frac{97011619 a_1^4}{174182400}-\frac{97011619 \pi ^2 
a_1^2}{174182400}+\frac{97011619 a_4}{7257600}+\frac{796232393699 \zeta
_3}{92990177280}-\frac{745372259 \pi 
^4}{4180377600} \\
&+\frac{5881974201847}{2092278988800}\Bigg) 
n_l^2+\Bigg(-\frac{19669747 \zeta_3}{21288960}-\frac{64 \pi 
^2}{4725}+\frac{3284183491}{2586608640}\Bigg) n_l^3 \,,
\end{split}
\end{equation}
\begin{equation}
\begin{split}
&\Pi^{(4)}_{\mathrm{le},3} =
\Bigg(-\frac{1875259367 a_1^4}{17740800}+\frac{1875259367 \pi ^2 
a_1^2}{17740800}-\frac{1024 \pi ^2 a_1}{2835}-\frac{1875259367 
a_4}{739200}\\
&-\frac{377287031234107 \zeta_3}{245887488000}+\frac{95566793477 
\pi ^4}{3193344000}+\frac{1175348 \pi 
^2}{1488375}-\frac{2388270016962373}{9958443264000}\Bigg) 
n_l^2 \\
&+\Bigg(\frac{12135758968 a_1^5}{172297125}-\frac{245658223193 
a_1^4}{6616209600}-\frac{12135758968 \pi ^2 
a_1^3}{103378275}+\frac{245658223193 \pi ^2 
a_1^2}{6616209600}\\
&-\frac{45718624634 \pi ^4 a_1}{516891375}-\frac{720316 
\pi ^2 a_1}{165375}-\frac{245658223193 a_4}{275675400}-\frac{97086071744 
a_5}{11486475}+\frac{8 \pi ^2 \zeta_3}{315}\\
&-\frac{1531450738927589 \zeta
_3}{3368252160000}+\frac{119140260224 \zeta
_5}{11486475}+\frac{176076905389817 \pi ^4}{31757806080000}+\frac{5337 \pi 
^2}{2450}\\
&+\frac{300002162759308069}{1500556337280000}\Bigg) n_l+\mathbf{si} 
\Bigg(\frac{22845879073 a_1^4}{6096384000}-\frac{22845879073 \pi ^2 
a_1^2}{6096384000}\\
&+\frac{22845879073 a_4}{254016000}+\frac{343009147408727 
\zeta_3}{6246309888000}-\frac{787819133821 \pi 
^4}{731566080000}+\frac{7118016595194017}{758926651392000}\Bigg) 
n_l^2\\
&+\Bigg(-\frac{7731286469 \zeta_3}{6227020800}-\frac{512 \pi 
^2}{42525}+\frac{81866930683}{50438868480}\Bigg) n_l^3 \,,
    \end{split}
  \end{equation}
}
where $n_l$ denotes contributions from closed lepton loops and where we
used the abbreviations $a_1 = \log 2$ and $ a_n =
\mathrm{Li}_n(1/2)$. $\mathbf{si}$ marks contributions from singlet
diagrams. These contributions are numerically tiny, amounting to at most
$5\%$ of each moment.

\subsection{High-energy expansion}
\label{sec:high-energy-expans}
In the high-energy region we write the result in the form
\begin{equation}
  \label{eq:7}
  \Pi^{(n)}_{\mathrm{he}} = \sum_{k=0} \Pi^{(n)}_{\mathrm{he},k} z^{-k}  \,.
\end{equation}
The coefficients of Eq.~(\ref{eq:7}) can be expressed through
four-loop massless propagator integrals which, in turn,
can be reduced to 28 master integrals.
This reduction has been done by evaluating sufficiently many
terms of the $1/D$ expansion
\cite{Baikov:2005nv}
of the corresponding coefficient functions
\cite{Baikov:1996rk}.
The master integrals are known analytically from
\cite{Baikov:2010hf} and numerically from~\cite{Smirnov:2010hd}.

As the result, the leading two coefficients of Eq.~(\ref{eq:7}) are
{\small
\begin{equation}
\begin{split}
& \Pi^{(4)}_{\mathrm{he},0}  =\mathbf{si}  \Bigg(\frac{73 a_1^4}{144}-\frac{73}{144} \pi ^2 
a_1^2+\frac{73 a_4}{6}-\frac{2 \zeta _3^2}{3}+\frac{5309 \zeta 
_3}{1120}+\frac{5 \zeta _5}{3}+\frac{2}{3} \zeta _3 \log (-4 
z) \\
&-\frac{11}{36} \log (-4 z)-\frac{2237 \pi 
^4}{17280}+\frac{1963}{3780}\Bigg)n_l^2 +n_l^2 \Bigg(\frac{53 
a_1^4}{60}-\frac{53}{60} \pi ^2 a_1^2+\frac{16 \pi ^2 a_1}{27}+\frac{106 
a_4}{5}-\zeta _3^2 \\
&+\frac{29129 \zeta _3}{1800}-\frac{125 \zeta 
_5}{18}-\frac{19}{12} \zeta _3 \log (-4 z)+\frac{5}{3} \zeta _5 \log (-4 
z)-\frac{1}{48} \log ^2(-4 z)-\frac{1}{12} \log (-4 z)\\
&-\frac{2161 \pi 
^4}{10800}-\frac{179 \pi ^2}{324}+\frac{3361}{900}\Bigg)+n_l 
\Bigg(-\frac{32 a_1^5}{225}+\frac{1559 a_1^4}{1080}+\frac{32}{135} \pi 
^2 a_1^3-\frac{1559 \pi ^2 a_1^2}{1080}\\
&+\frac{106 \pi ^4 
a_1}{675}+\frac{59 \pi ^2 a_1}{12}+\frac{1559 a_4}{45}+\frac{256 
a_5}{15}-\frac{\pi ^2 \zeta _3}{24}+\frac{6559 \zeta _3}{320}-\frac{1603 
\zeta _5}{120}-\frac{35 \zeta _7}{4}\\
&+\frac{23}{128} \log (-4 z)-\frac{59801 
\pi ^4}{129600}-\frac{157 \pi ^2}{72}-\frac{71189}{34560}\Bigg)+n_l^3 
\Bigg(-\frac{15109 \zeta _3}{22680}-\frac{5 \zeta _5}{9}\\
&-\frac{1}{9} \zeta 
_3 \log ^2(-4 z)+\frac{19}{27} \zeta _3 \log (-4 z)-\frac{1}{108} \log ^3(-4 
z)+\frac{11}{72} \log ^2(-4 z)-\frac{151}{162} \log (-4 z)\\
&+\frac{8 \pi 
^2}{405}+\frac{75259}{68040}\Bigg) \,,
\end{split}
\end{equation}
\begin{equation}
\begin{split}
& \Pi^{(4)}_{\mathrm{he},1}  =n_l^2 \Bigg(\frac{a_1^4}{9}-\frac{1}{9} \pi ^2 a_1^2+\frac{8 \pi ^2 
a_1}{9}+\frac{8 a_4}{3}-\frac{\zeta _3^2}{6}+\frac{3023 \zeta 
_3}{432}-\frac{415 \zeta _5}{54}\\
&-\frac{167}{72} \zeta _3 \log (-4 
z)-\frac{35}{36} \zeta _5 \log (-4 z)+\frac{3}{16} \log ^3(-4 
z)-\frac{15}{16} \log ^2(-4 z)+\frac{1}{4} \pi ^2 \log (-4 
z)\\
&+\frac{539}{144} \log (-4 z)+\frac{5 \pi ^4}{108}-\frac{103 \pi 
^2}{108}-\frac{491}{216}\Bigg)+n_l \Bigg(-\frac{3}{4} \pi ^2 a_1 \log (-4 
z)-\frac{a_1^4}{2}+\frac{1}{2} \pi ^2 a_1^2\\
&+\frac{31 \pi ^2 a_1}{4}-12 
a_4-2 \zeta _3^2-\frac{\pi ^2 \zeta _3}{16}-\frac{3647 \zeta 
_3}{192}-\frac{1825 \zeta _5}{192}+\frac{10073 \zeta _7}{384}+\frac{41}{32} 
\zeta _3 \log (-4 z)\\
&+\frac{35}{16} \zeta _5 \log (-4 z)-\frac{9}{32} \log 
^3(-4 z)+\frac{9}{64} \log ^2(-4 z)+\frac{15}{32} \pi ^2 \log (-4 
z)-\frac{685}{128} \log (-4 z)\\
&-\frac{\pi ^4}{48}-\frac{673 \pi 
^2}{192}-\frac{47}{36}\Bigg)+\mathbf{si}\Bigg(-\frac{9 \zeta _3^2}{2}-\frac{13 \zeta 
_3}{3}-\frac{5 \zeta _5}{12}+\frac{147 \zeta _7}{16}+\frac{1}{3}\Bigg) 
 n_l^2\\
&+n_l^3 \Bigg(-\frac{4 \zeta _3}{9}+\frac{2}{3} \zeta _3 
\log (-4 z)-\frac{1}{36} \log ^3(-4 z)+\frac{13}{72} \log ^2(-4 
z)-\frac{317}{216} \log (-4 z)\\
&+\frac{4 \pi ^2}{135}+\frac{25}{27}\Bigg) \,,
\end{split}
\end{equation}
}
where $n_l$ denotes contributions from closed lepton loops and where we
used the abbreviations $a_1 = \log 2$ and $ a_n =
\mathrm{Li}_n(1/2)$. $\mathbf{si}$ marks contributions from singlet
diagrams. Contrary to the low-energy case these are sizeable and
comparable to the contributions from other diagram classes. Naturally
the same holds true for the contributions to the anomalous magnetic
moment, which are dominated by the high-energy region (see
Section~\ref{sec:muon}).

\subsection{Threshold expansion}
\label{sec:threshold-expansion}
The polarization function in the threshold region can be written as
\begin{equation}
  \label{eq:pi_thr}
  \Pi^{(n)}_{\mathrm{thr}} = 16\*\pi^2\*\sum_{k=2-n} \Pi^{(n)}_{\mathrm{thr},k}
  \, \big(\sqrt{1-z}\,\big)^{k}  \,.
\end{equation}
In non-relativistic quantum field theory the NNLO expression for the 
threshold cross section is known for an arbitrary $SU(N)$ gauge
group~\cite{Hoang:2000yr}. This means that the derivation of the threshold expansion
is essentially the same as for the QCD case, which is described in 
Refs.~\cite{Hoang:2008qy,Kiyo:2009gb}. The only additional steps consist
of setting the group invariants to their $U(1)$ values and converting
the coupling to the on-shell scheme.

The resulting coefficients read
\begin{align}
  \label{pi3_thr_coeff}
  \Pi^{(3)}_{\text{thr},-1} =& \frac{\pi^3}{384}\*n_l\,,\\  
\label{pi3_thr_coeff_0}
  \Pi^{(3)}_{\text{thr},0} =& \bigg(\frac{17}{768}+\frac{11}{128}\*a_1\bigg)\*n_l\*\log(1-z)+\frac{11}{512}\*n_l\*\log^2(1-z)+\text{const}\,,\\
\Pi^{(3)}_{\text{thr},1} =
&\bigg(\frac{17}{192\*\pi}-\frac{397}{2304}\*\pi+\frac{5}{24}\*\pi\*a_1-\frac{13}{2304}\*\pi^3+\frac{41}{64\*\pi}\*\zeta_3\bigg)\*n_l\notag\\
&+\frac{11}{192}\*\pi\*n_l\*\log(1-z)
+\bigg(-\frac{11}{72\*\pi}+\frac{\pi}{72}\bigg)\*n_l^2\,,
\end{align}
for $n=3$ and
\begin{align}
  \label{pi4_thr_coeff_-2}
  \Pi^{(4)}_{\text{thr},-2} =& \frac{1}{128}\*\pi^2\*\zeta_3\*n_l\,,\\
  \label{pi4_thr_coeff_-1}
  \Pi^{(4)}_{\text{thr},-1} =&
  \bigg(\frac{7}{2304}\*\pi^3-\frac{11}{384}\*\pi^3\*a_1-\frac{11}{64}\*\pi\*\zeta_3\bigg)\*n_l-\frac{11}{768}\*\pi^3\*n_l\*\log(1-z)\,,\\
  \label{pi4_thr_coeff_0}
  \Pi^{(4)}_{\text{thr},0} =&
\bigg(-\frac{467}{18432}+\frac{115}{384}\*a_1-\frac{121}{256}\*a_1^2-\frac{2207}{9216}\*\pi^2+\frac{5}{48}\*\pi^2\*a_1\notag\\
&+\frac{9}{4096}\*\pi^4+\frac{131\*\zeta_3}{512}\bigg)\*n_l\*\log(1-z)\notag\\
&+\bigg(-\frac{839}{9216}+\frac{\pi^2}{144}\bigg)\*n_l^2\*\log(1-z)\notag\\
&+\bigg(\frac{115}{1536}-\frac{121}{512}\*a_1+\frac{11}{768}\*\pi^2\bigg)\*n_l\*\log^2(1-z)\notag\\
&-\frac{121}{3072}\*n_l\*\log^3(1-z)+\text{const}\,,
\end{align}
for $n=4$, where we again used $a_1 = \log 2$. The non-logarithmic
contributions to $\Pi^{(3)}_{\text{thr},0}$ and
$\Pi^{(4)}_{\text{thr},0}$, denoted here by ``const'', are not available
in the literature.

\subsection{Pad\'e approximation}
\label{sec:pade-approximation}
Having all building blocks at hand the polarization function can be
reconstructed using Pad\'e
approximation~\cite{Baker:1975,Broadhurst:1993mw,Fleischer:1994ef,Broadhurst:1994qj,Baikov:1995ui,Chetyrkin:1995ii,Chetyrkin:1996cf,Hoang:2008qy,Kiyo:2009gb}. For
the four-loop contribution we closely follow the procedure outlined in
Ref.~\cite{Kiyo:2009gb}. In the three-loop case we introduce slight
modifications to accommodate the large amount of information in the
available low- and high-energy expansions. We first give a brief review
of the method as used in the four-loop case and then discuss the changes
for the application to the three-loop contribution in
Section~\ref{sec:pade-approximation-3}.

\subsubsection{Approximation procedure at four loops}
\label{sec:pade-approximation-4}
As in Ref.~\cite{Kiyo:2009gb}, we first split $\Pi^{(4)}(z)$ into two parts,
\begin{equation}
  \label{eq:pi_split}
  \Pi^{(4)}(z) = \Pi^{(4)}_{\text{reg}}(z) + \Pi^{(4)}_{\text{log}}(z)\,,
\end{equation}
using the ansatz
\begin{align}
\label{eq:pi4_log}
  \Pi^{(4)}_{\text{log}}(z) =& \sum_{j=-1}^0\sum_{i=1}^{3+2\*j} k_{ij} {\cal K}^{(4)}_{ij}(z) +
  \sum_{n=0}^{1}\sum_{m=0}^{3} d_{mn} {\cal D}^{(4)}_{mn}(z)\,,\\
\label{eq:pi4_log_thr}
  {\cal K}^{(4)}_{ij}(z) =& \Pi^{(2)}(z)^i\*G(z)^j \times
  \begin{cases}
    \left(A_1+\frac{1}{z}\right)&\text{for }i=3,j=0\\
    \left(A_2+\frac{1}{z}\right)& \text{for }i=1,j=-1\\
    \quad1&\text{otherwise}
  \end{cases}\,,
\\
\label{eq:pi4_log_he}
  {\cal D}^{(4)}_{mn}(z) =&
  \big(z\,\*G(z)\big)^m\*\left(1-1/z\right)^{\lceil\frac{m}{2}\rceil}\frac{1}{z^n}\times
  \begin{cases}
    \left(1+\frac{1}{B_1z}\right)& \text{for }n=3\\
    \left(1+\frac{1}{B_2z}\right)& \text{for }n=2\\
    \quad1&\text{otherwise}
  \end{cases}\,,
\end{align}
with 
%arbitrary auxiliary parameters $A_1, A_2, B_1, B_2$, 
the known two-loop polarization function $\Pi^{(2)}(z)$ as
introduced in Eq.~(\ref{eq:5}) and
\begin{equation}
  \label{eq:G_def}
  G(z) = \frac{1}{2z}\frac{\log\left(u\right)}{\sqrt{1-\frac{1}{z}}}\,,
\qquad u = \frac{\sqrt{1-\frac{1}{z}}-1}{\sqrt{1-\frac{1}{z}}+1}\,.
\end{equation}

The coefficients $k_{ij}$ can be fixed so that all known
logarithms in the threshold expansion
(Eqs.~\eqref{eq:pi_thr},~\eqref{pi4_thr_coeff_-2}--~\eqref{pi4_thr_coeff_0})
 are absorbed into the first term on the right-hand side of
 Eq.~\eqref{eq:pi4_log}. Expanding Eq.~\eqref{eq:pi4_log_thr} in the
 threshold region generates the required logarithms and half-integer
 powers of $1-z$, as can be seen from the expansions
\begin{align}
  \label{eq:pi_G_thr}
  G(z) &= \frac{\pi}{2}\frac{1}{\sqrt{1-z}}+{\cal O}((1-z)^0)\,,\\
\Pi^{(2)}(z)=&-\frac{1}{16}\log\left(1-z\right)+\text{const} +{\cal O}(\sqrt{1-z}) \,.
\end{align}

In a similar way the factor $z\,\*G(z)$ in Eq.~\eqref{eq:pi4_log_he}
generates logarithms in the high-energy region:
\begin{equation}
  \label{G_he}
  z\,\*G(z) = -\frac{1}{2}\log(-4\*z)+{\cal O}\left(\frac{1}{z}\right)\,.
\end{equation}
This means we can choose the coefficients $d_{mn}$ with $m>0$ in such a way that
these logarithms are also absorbed into $\Pi^{(4)}_{\text{log}}(z)$. The
remaining coefficients $d_{0n}$ are fixed by requiring that
$\Pi^{(4)}_{\text{reg}}(z)$ has no poles for $z\to 0$ and, more specifically,
$\Pi^{(4)}_{\text{reg}}(0) = 0$.

To estimate the error of the
approximation we vary the parameters $A_i,B_i$ independently with
\begin{align}
  \label{eq:pade_params}
  A_i \in & \{-1 \pm 1,\, -1\pm 4,\, -1 \pm 16,\, -1 \pm 64\}\,,\notag\\
  B_i \in & \{\pm1,\, \pm 4,\, \pm 16,\, \pm64\}\,.
\end{align}

In a second step we define
\begin{equation}
  \label{eq:pade_aux_4l}
  P_1(\omega) =
  \left(\frac{1-\omega}{1+\omega}\right)^2\*\left(\Pi^{(4)}_{\text{reg}}\big(z(w)\big)
    - \Pi^{(4)}_{\text{reg}}(-\infty)\right)
\end{equation}
with $z(\omega) = 4\*\omega/(1+\omega)^2$ and construct Pad\'e
approximants
\begin{equation}
  \label{eq:pade_def}
  p_{n,m}(\omega) =  \frac{\sum_{i=0}^{n-1} a_i \omega^i+\omega^n}{\sum_{i=0}^m b_i \omega^i}
\end{equation}
using the constraints
\begin{align}
  \label{eq:pade_cons}
  p_{n,m}(1)=&P_1(1), & p'_{n,m}(1)=&P'_1(1),\notag\\
  p_{n,m}(-1)=&P_1(-1), & p^{(i)}_{n,m}(0)=&P^{(i)}_1(0),& &i=0,\dots,3 \,,
\end{align}
and requiring the absence of terms proportional to $z^{-1/2}$ in the
high-energy expansion of $\Pi^{(4)}_{\text{reg}}$. In total there are
eight constraints, i.e. we obtain Pad\'e approximants with $n+m=7$.

If $p_{n,m}(\omega)$ has poles inside the unit circle $|\omega|<1$ the
corresponding reconstructed polarization function shows unphysical
singularities in the complex $z$ plane. We therefore discard such
approximants. Furthermore, we require $|p_{n,m}(\omega)| < 1.5$ for
$|\omega|=1$ in order to remove approximants with pronounced additional
peaks above the physical threshold at $z=1$. A notable effect of this
cut is the elimination of all Taylor approximants $p_{7,0}(\omega)$.

\begin{figure}
  \centering
  \includegraphics[width=0.8\textwidth]{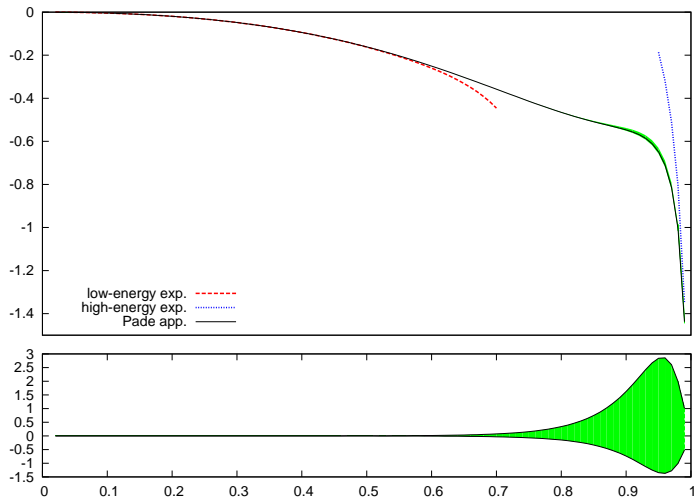}
  \caption{Pad\'e approximation for $\Pi^{(4)}\left (-\frac{x^2}{1-x}m_\mu^2\right )$ with muon loops. We show in the top
    half the approximants and in the bottom half the relative error
    with respect to the local mean of all approximants
    obtained.}
  \label{fig:app}
\end{figure}

Since the approximants are only available in
numerical form we show the general features in Fig.~\ref{fig:app}.
At the top of the figure we show the envelope of all Pad\'e approximants
that have been calculated and compare with the low- and high-energy
expansions. In the bottom half we show the maximal deviation from the
mean in percent. As can be seen there are only significant deviations of
the order of a few percent in the range $x>0.8$, which correspond to the
uncertainty due to a limited number of terms in the high-energy
expansion.

\subsubsection{Modifications at three loops}
\label{sec:pade-approximation-3}
At three loops much deeper expansions can be used for the construction
of the Pad\'e approximants. The 30 known coefficients in the low-energy
expansion~\cite{Boughezal:2006uu,Maier:2007yn} together with 31
coefficients from the high-energy expansion~\cite{Maier:2011jd} and
three coefficients in the threshold expansion lead to a significantly
larger system of constraints compared to the four-loop case.

Our strategy will be to incorporate as much of the threshold information
as possible into the logarithmic function $\Pi^{(3)}_{\text{log}}(z)$,
so that all constraints are imposed at $\omega=0$. The resulting system
can be solved very efficiently using well-established techniques of
one-point Pad\'e approximation\footnote{One-point approximation means
that the value of the approximated function and its derivatives at a
single point are known.}.

As in the four-loop case we first split the polarization function into
two parts, using
\begin{align}
  \label{eq:pi3_log}
  \Pi^{(3)}_{\text{log}}(z) =& \sum_{i\geq0,j} k_{ij} {\cal K}^{(3)}_{ij}(z) +
  \sum_{m,n} d_{mn} {\cal D}^{(3)}_{mn}(z) \,,\\
  {\cal K}^{(3)}_{ij}(z) =& \Pi^{(2)}(z)^i\*G(z)^j \times
  \begin{cases}
    (1-z)^j & \text{for }j>0\\
    \quad 1 & \text{otherwise}
  \end{cases}\,,
\\
  {\cal D}^{(3)}_{ij}(z) =&
  \big(z\,\*G(z)\big)^m\*\left(1-1/z\right)^{\lceil\frac{m}{2}\rceil}\frac{1}{z^n}
\end{align}
to absorb the logarithms and the threshold singularity. Since we expect
to obtain sufficiently many different Pad\'e approximants for a reliable error
estimate we refrain from introducing additional parameters at this point.

To map all available information from the low- and high-energy expansions onto
$\omega=0$, we define
\begin{align}
  \label{eq:pade_aux_3l}
  P_{30}(\omega) &= z(\omega)^{31}\*\left(\Pi^{(3)}_{\text{reg}}\big(z(\omega)\big)
    - \sum_{i=0}^{30}\frac{H^{(3)}_i}{z(\omega)^i}\right)\,,\notag\\
 H^{(3)}_i &= \frac{1}{i!}\left(\frac{\partial}{\partial(1/z)}\right)^i\Pi^{(3)}_{\text{reg}}(z)\Bigg|_{z\to -\infty}\,.
\end{align}
The Taylor approximant is fixed by imposing
\begin{equation}
  \label{eq:pade_taylor}
  p^{(i)}_{61,0}(0) = P^{(i)}_{30}(0)\,,\qquad i=0,\dots,61\,.
\end{equation}
We can deduce all further approximants with $n+m=61$ and $n+m=60$
using Baker's recursion formula~\cite{baker:1970}\footnote{Note that
the denominator in the first equation of Eq.~(13) 
in Ref.~\cite{baker:1970} contains an obvious typo. The proper expression is
$\bar{\eta}_{2j-1}\*\theta_{2j-2}(x) - x\*\bar{\eta}_{2j-2}\*\theta_{2j-1}(x)$.
}
\begin{align}
  \label{eq:pade_rec}
p_{n-\frac{j}{2},\frac{j}{2}} 
&= \frac{\hat{\eta}_j}{\hat\theta_j}
= \frac{\hat\eta_{j-2} -
  \omega\*\,\hat\eta_{j-1}}{\hat\theta_{j-2} -
  \omega\*\,\hat\theta_{j-1}} & j \text{ even}\,,\\
p_{n-\frac{j+1}{2},\frac{j-1}{2}} 
&= \frac{\hat{\eta}_j}{\hat\theta_j}
= \frac{\hat\eta_{j-2} -
  \hat\eta_{j-1}}{\hat\theta_{j-2} -
  \hat\theta_{j-1}} & j \text{ odd}\,,
\end{align}
where the numerator $\hat{\eta}_j$ of the Pad\'e
approximant is normalised as in Eq.~\eqref{eq:pade_def}.

It should be noted that the known threshold constant at order $1-z$ is
not included in the approximants. We find however that the information
from the threshold region has virtually no effect on the reconstructed
polarization functions in the Euclidean regime $z<0$.\footnote{If not
all constraints from the low- and high-energy regions are taken into
account, the importance of the threshold information increases.  Still,
not taking into account any threshold coefficients changes the
polarization function by less than one per mille in the euclidean
region, even if only three terms each from the low- and the high-energy
expansion are taken into account.} This is contrary to the four-loop
case, where the approximants are far less constrained by the low- and
high-energy expansions.

\section{The anomalous magnetic moment of the muon at five loops}
\label{sec:muon}
The QED corrections to the anomalous magnetic moment can be calculated
in perturbation theory and can thus be written in form of a power series
in the fine structure constant $\alpha$
\begin{equation}
  a_\mu = \sum_{k=1}^\infty \left ( \frac{\alpha}{\pi}\right )^k a_\mu
  ^{(2 k)} \,,
\end{equation}
where $a_\mu^{(2 k)}$ can be further decomposed -- following the
conventions in Ref.~\cite{Aoyama:2012wk} -- as 
\begin{equation}
  a_\mu^{(2 k)} = A_1^{(2 k)}  
  + A_2^{(2 k)}(m_e / m_\mu) + A_2^{(2 k)}(m_\tau / m_\mu)  
  + A_3^{(2 k)}(m_e / m_\mu,m_\tau / m_\mu) \,.
\end{equation}
$A_1^{(2 k)}$ contains the universal contributions, which in case of
the muon anomalous magnetic moment only contain muon loops. The
diagrams contributing to $A_2^{(2 k)}(m_e / m_\mu)$ and $A_2^{(2
  k)}(m_\tau / m_\mu)$ have at least one electron or tau loop,
respectively. In $A_3^{(2 k)}(m_e / m_\mu,m_\tau / m_\mu)$
contributions from diagrams with both electron and tau loops are
collected. In this paper we are mainly interested in contributions to 
$A_2^{(2 k)}(m_e / m_\mu)$ without any muon loops. 

\begin{figure}
  \centering
  \includegraphics[width=0.3\linewidth]{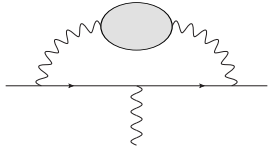}
  \caption{Prototype diagram}
  \label{fig:1}
\end{figure}

\begin{figure}
  \centering
  I(a)\includegraphics[width=0.15\linewidth]{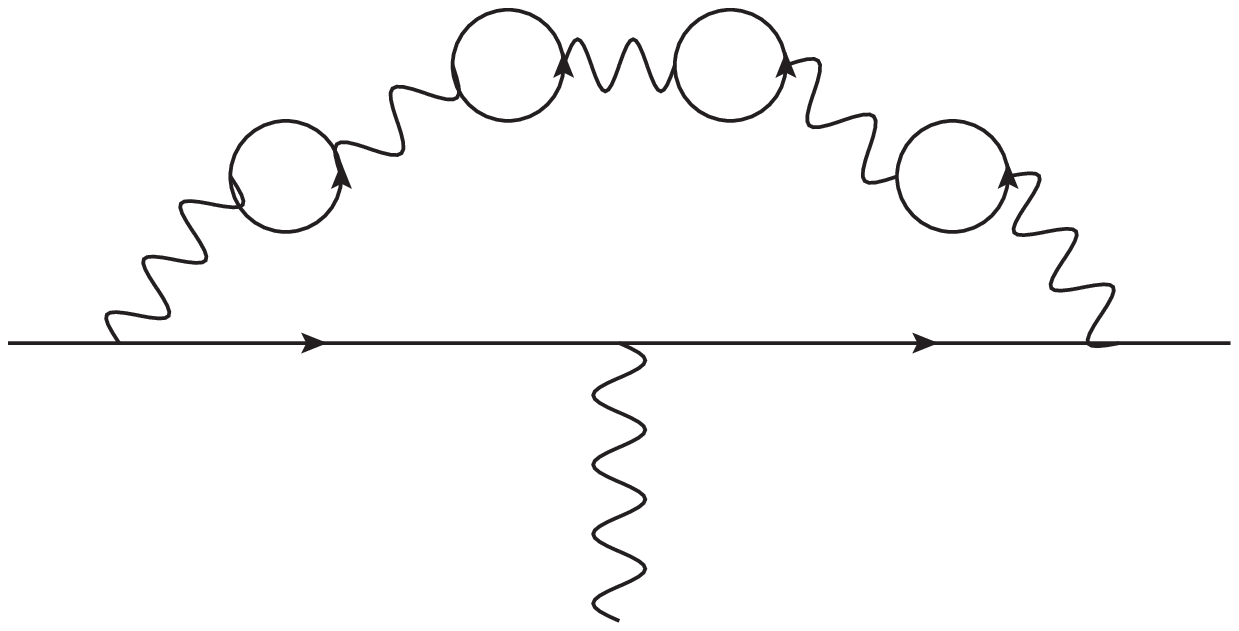}
  I(b)\includegraphics[width=0.15\linewidth]{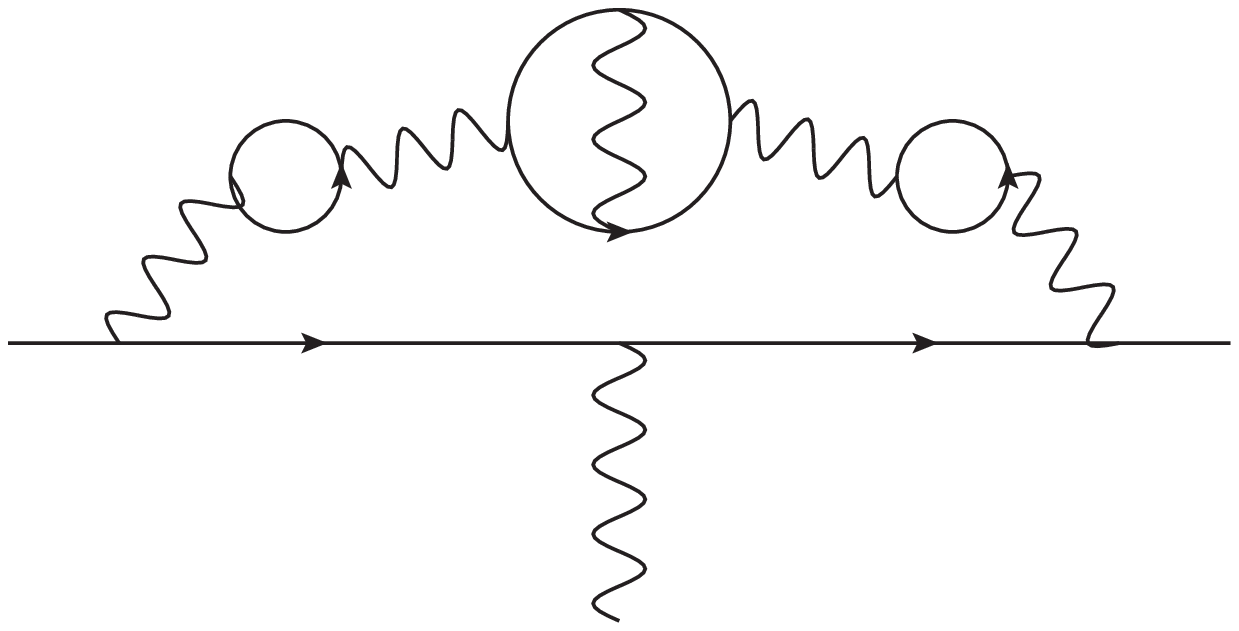}
  I(c)\includegraphics[width=0.15\linewidth]{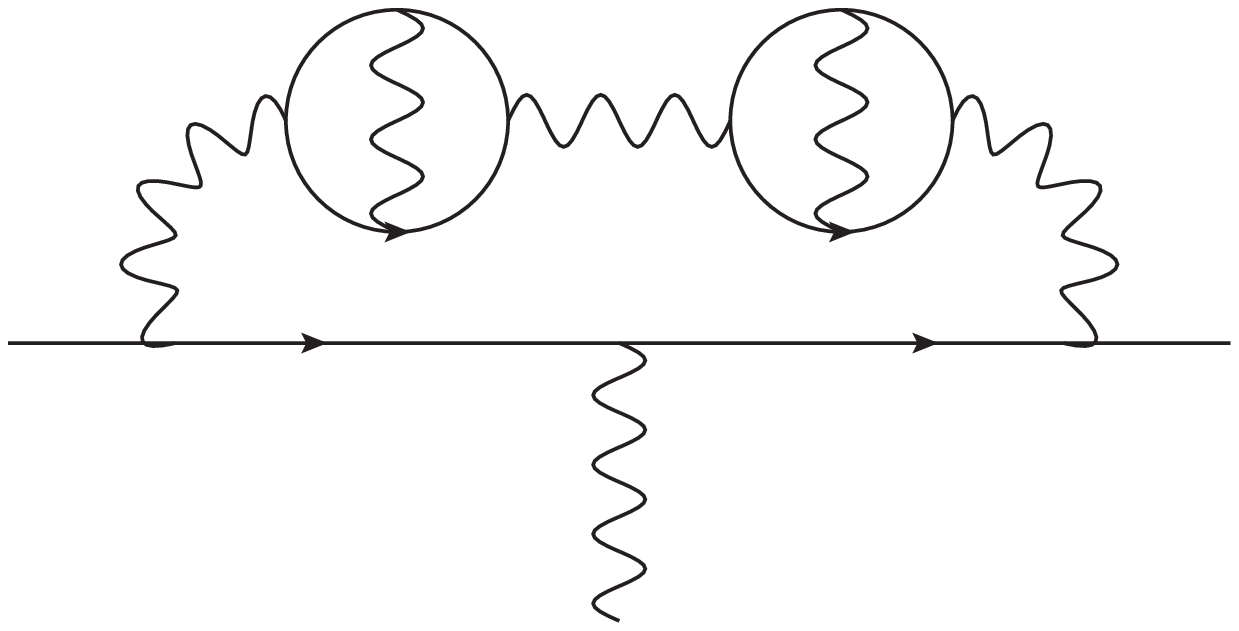}
  I(d)\includegraphics[width=0.15\linewidth]{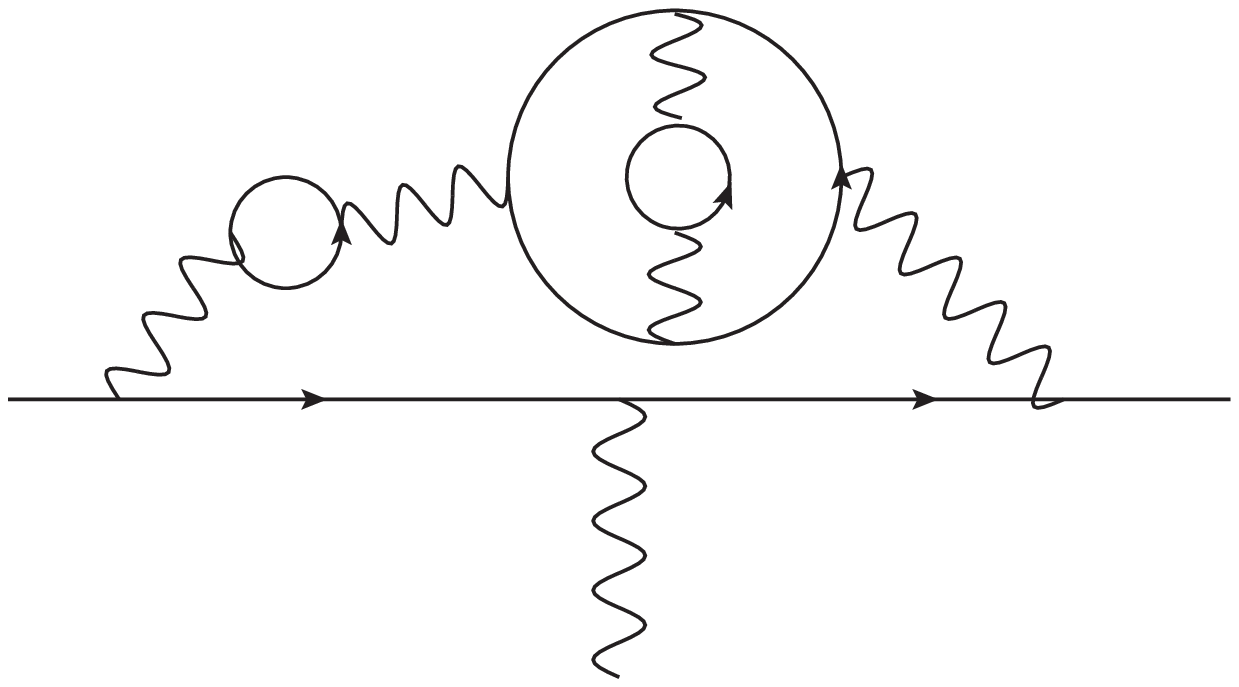}
  I(e)\includegraphics[width=0.15\linewidth]{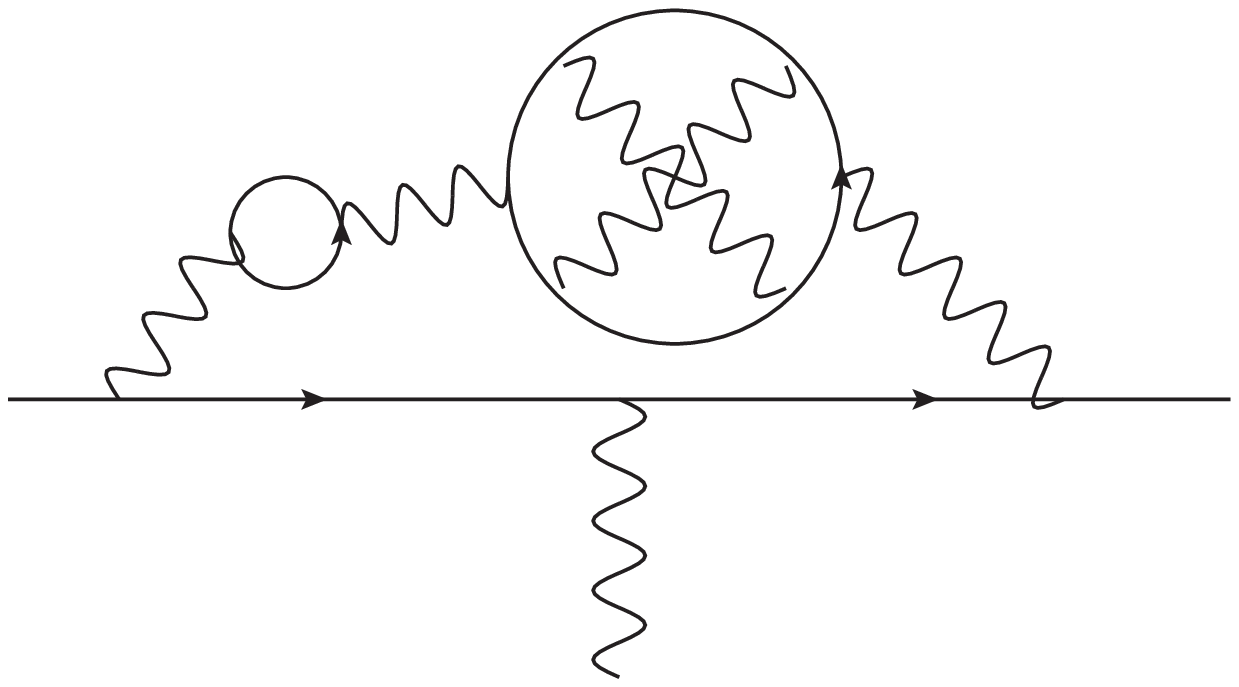}
  I(f)\includegraphics[width=0.15\linewidth]{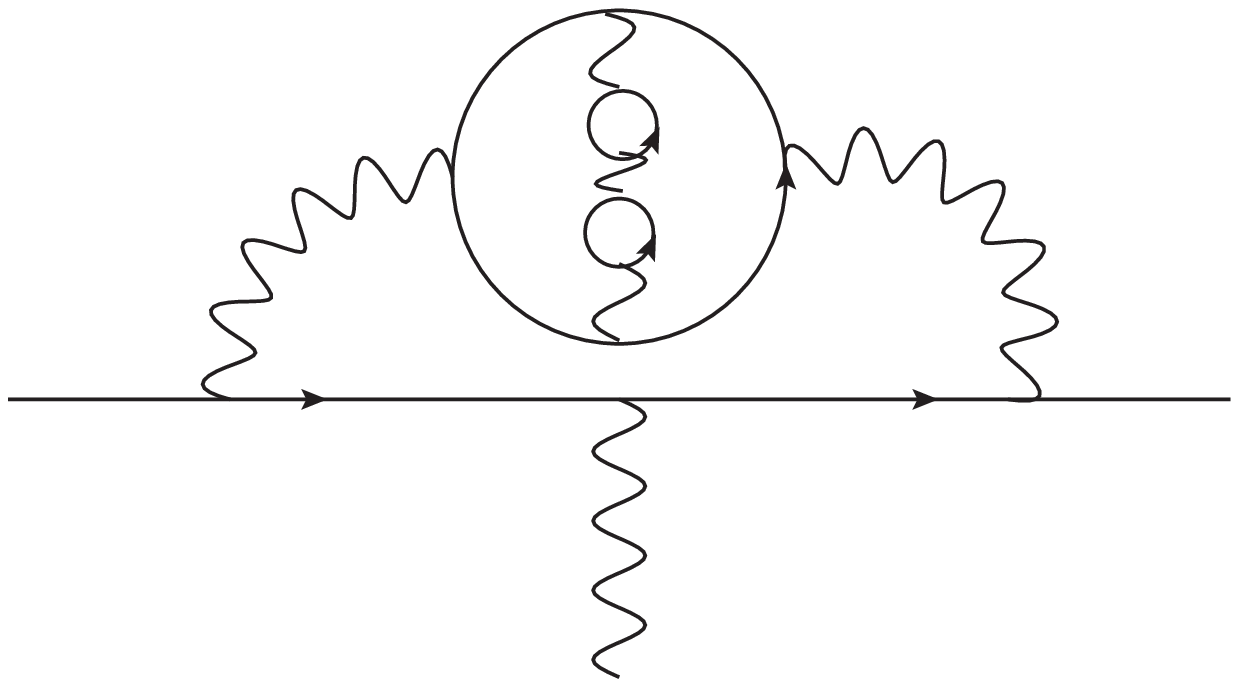}
  I(g)\includegraphics[width=0.15\linewidth]{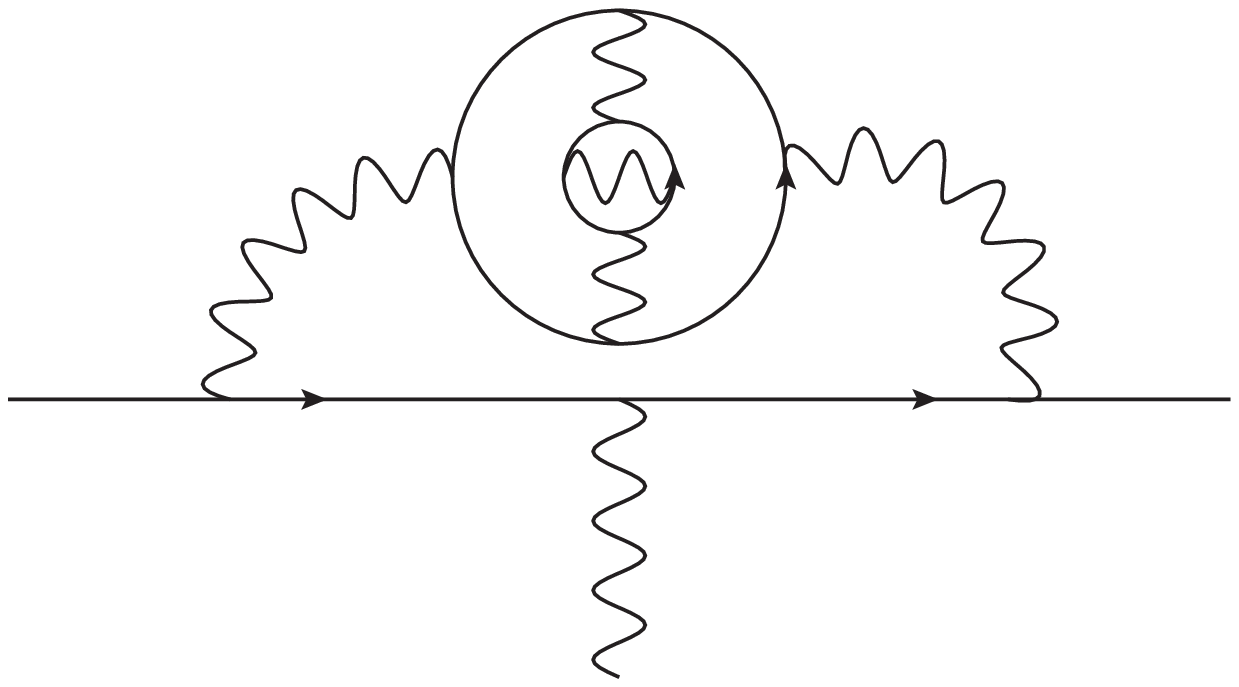}
  I(h)\includegraphics[width=0.15\linewidth]{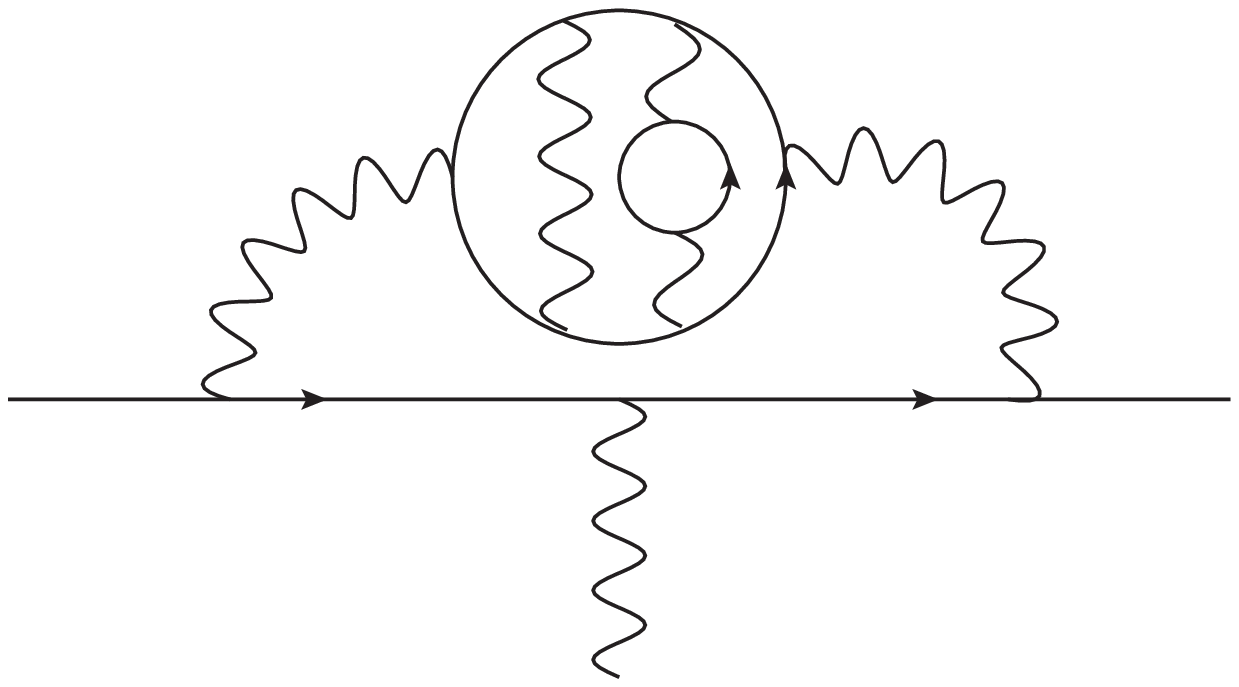}
  I(i)\includegraphics[width=0.15\linewidth]{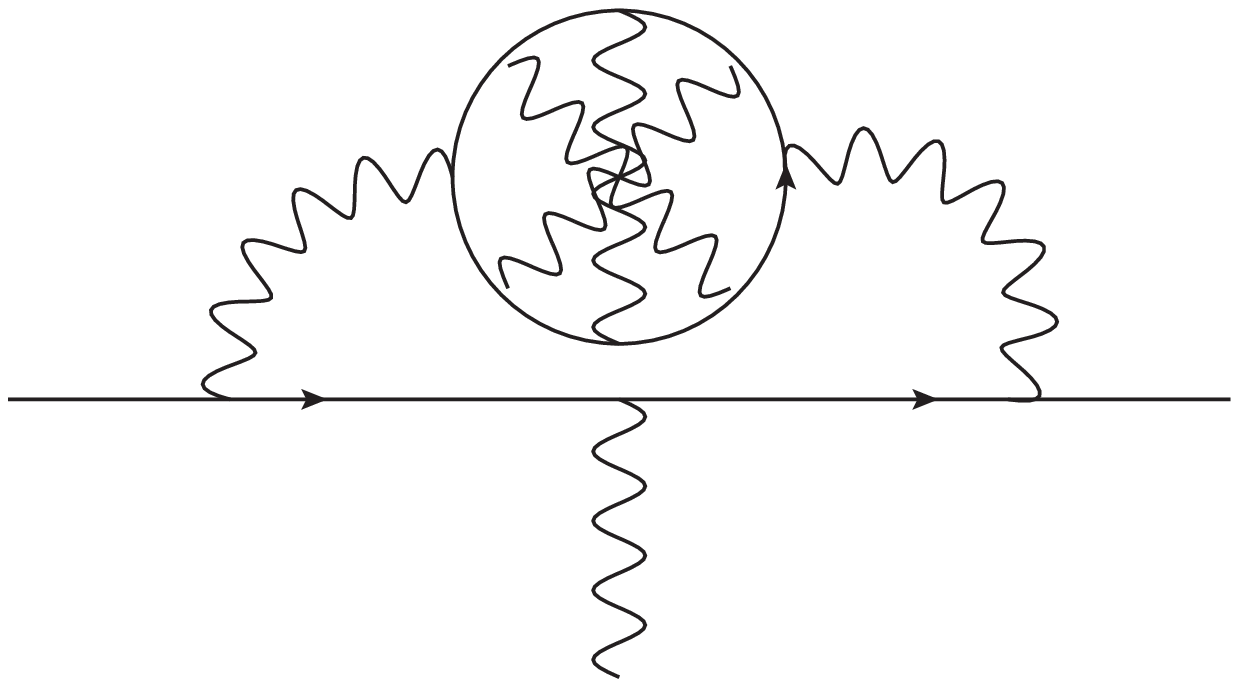}
  I(j)\includegraphics[width=0.15\linewidth]{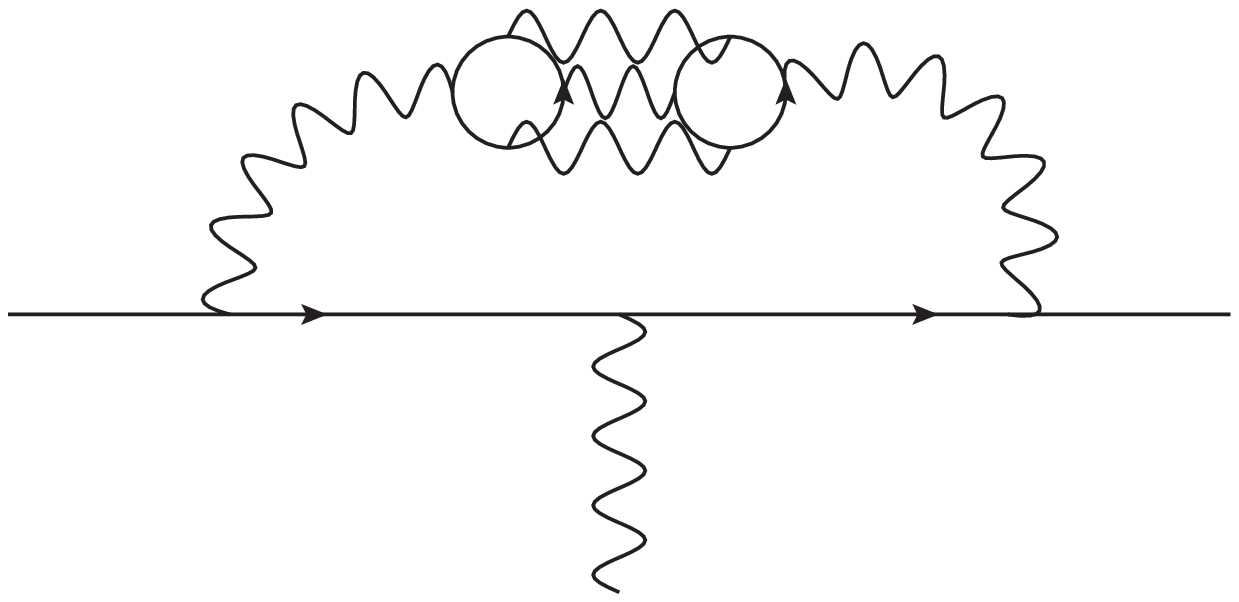}
  \caption{Classes of diagrams accessible by the used method.}
  \label{fig:2}
\end{figure}

{
The contributions to the anomalous magnetic moment of the
muon due to photon polarization effects can be calculated
(cf. Fig.~\ref{fig:1}) by using~\cite{Lautrup:1974ic}
% \begin{equation}
% \label{eq:1}
% a_\mu = \frac{\alpha}{\pi} \int_{0}^1 \dd x (1-x)  \left [ - \Pi
%   (s_x) \right ] \,, \, s_x = - \frac{x^2}{1-x} m_\mu^2 \,.
% \end{equation}
\begin{equation}
\label{eq:1new}
a_\mu = \frac{\alpha}{\pi} \int_{0}^1 \dd x (1-x)  \frac{1}{1 + \Pi
  (s_x) } \,, \, s_x = - \frac{x^2}{1-x} m_\mu^2 \,.
\end{equation}
This formula can be obtained by considering the one-loop result for
$g-2$ for the case of a heavy photon in combination with the
dispersion relation for $\Pi(q^2)$.
% {\color[gray]{.5}
% Multiple insertions of one-particle irreducible vacuum polarization can
% be obtained by using
% \begin{equation}
% \label{eq:2}
% a_\mu = \frac{\alpha}{\pi} \int_{0}^1 \dd x (1-x)  \left [ - \Pi^{\mathrm{1PI}}
%   (s_x) \right ]^n \,,
% \end{equation}
% }
The right-hand side of Eq.~(\ref{eq:1new}) should be expanded in
  $\alpha$, 
leading to e.g. at two-loop order
% \begin{equation}
% \label{eq:3}
% a_\mu = \frac{\alpha}{\pi} \int_{0}^1 \dd x (1-x)  \left [ -
%   \Pi^{\mathrm{2L}}
%   (s_x) + 2 \Pi^{\mathrm{1L}}
%   (s_x) \Pi^{\mathrm{1L}}
%   (s_x)\right ] \,.
% \end{equation}
\begin{equation}
\label{eq:3new}
a_\mu^{(6)} = \int_{0}^1 \dd x (1-x)  \left [ -
  \Pi^{(2)}(s_x) +  \big(-\Pi^{(1)}(s_x)\big)^2\right ] \,.
\end{equation}
}
The classes of diagrams accessible by this method are shown in
Fig.~\ref{fig:2}. 
\newcommand{\zwf}[3]{\ensuremath{ #1 {{\scriptstyle + #2} \atop
      {\scriptstyle - #3} } }}

\begin{table}
  \centering
\begin{tabular}{l||l|l|ll||}
  &this work&Ref.~\cite{Baikov:2012rr}&
  \multicolumn{2}{|l||}{Refs.~\cite{Kinoshita:2005sm,Aoyama:2008hz,Aoyama:2010zp,Aoyama:2008gy}}\\\hline\hline
  I(a)&20.142 813&20.183 2&20.142 93(23)&\cite{Kinoshita:2005sm}\\\hline
  I(b)&27.690 061&27.718 8&27.690 38(30)&\cite{Kinoshita:2005sm}\\\hline
  I(c) &\phantom{0}4.742 149&\phantom{0}4.817 59&\phantom{0}4.742 12(14)&\cite{Kinoshita:2005sm}\\\hline
  I(d)+I(e)&\phantom{0}6.241 470&\phantom{0}6.117 77&\phantom{0}6.243 32(101)(70)&\cite{Kinoshita:2005sm}\\\hline
  I(e)&-1.211 249&-1.331 41&-1.208 41(70)&\cite{Kinoshita:2005sm}\\\hline
  I(f)+I(g)+I(h)&\zwf{\mbox{\phantom{0}4.446 8}}{6}{4}&\phantom{0}4.391 31&\phantom{0}4.446 68(9)(23)(59)&\cite{Kinoshita:2005sm,Aoyama:2008hz}\\\hline
  I(i)&\zwf{\mbox{\phantom{0}0.074 6}}{8}{19}&\phantom{0}0.252 37&\phantom{0}0.087 1(59)&\cite{Aoyama:2010zp}\\\hline
  I(j)&\zwf{\mbox{-1.246 9}}{4}{3}&-1.214 29&-1.247 26(12)&\cite{Aoyama:2008gy}
\end{tabular}
%   \begin{tabular}{l||l|l|l||}
%     &this work&Ref.~\cite{Baikov:2012rr}&Refs. cited in ~\cite{Baikov:2012rr}\\\hline\hline
% I(a)&20.142 813&20.183 2&20.142 93(23)\\\hline
% I(b)&27.690 061&27.718 8&27.690 38(30)\\\hline
% I(c) &4.742 149&4.817 59&4.742 12(14)\\\hline
% I(d)+I(e)&6.241 470&6.117 77&6.243 32\\\hline
% I(e)&-1.211 249&-1.331 41&-1.208 41(70)\\\hline
% I(f)+I(g)+I(h)&\zwf{\mbox{4.446 9}}{6}{4}&4.391 31&4.446 68\\\hline
% I(i)&\zwf{\mbox{0.074 6}}{8}{19}&0.252 37&0.087 1(59)\\\hline
% I(j)&\zwf{\mbox{-1.246 9}}{4}{3}&-1.214 29&-1.247 26(12)
%   \end{tabular}
  \caption{Results for $A_2^{(10)}(m_e/m_\mu)$ with pure electronic
insertions. The errors listed in the second column are estimated from
the spread between different Pad\'e approximants, which is negligible
for classes I(a)--I(e). Please note that the authors of
Ref.~\cite{Baikov:2012rr} only used the asymptotic form of $\Pi(s)$ and
did not provide any error estimate.}
  \label{tab:Table1}
\end{table}
\begin{table}
  \centering
  \begin{tabular}{c||l|l||}
    &this work&Ref.~\cite{Aoyama:2012wk}\\\hline
I(a)&22.566 976&22.566 973 (3)
\\\hline
I(b)&30.667 093&30.667 091 (3)\\\hline
I(c)&\phantom{0}5.141 395&\phantom{0}5.141 395 (1)\\\hline
I(e)&-0.931 839& -0.931 2 (24)
  \end{tabular}
  \caption{Results for $A_2^{(10)}(m_e/m_\mu)$ including electronic and
    muonic contributions.}
  \label{tab:Table2}
\end{table}
\begin{table}
  \centering
  \begin{tabular}{l||l|l||}
    &this work&Ref.~\cite{Aoyama:2012wj}\\\hline\hline
I(a)&0.000 471&0.000 470 94 (6)
\\\hline
I(b)&0.007 010&0.007 010 8 (7)
\\\hline
I(c) &0.023 467&0.023 468 (2)
\\\hline
I(d)+I(e)&0.014 094&0.014 098(5)(4)
\\\hline
I(e)&0.010 291&0.010 296 (4)
\\\hline
I(f)+I(g)+I(h)&\zwf{\mbox{0.037 85}}{5}{3} &0.037 833(20)(6)(13)
\\\hline
I(i)&\zwf{\mbox{0.017 21}}{8}{23}&0.017 47 (11)
\\\hline
I(j)&\zwf{\mbox{0.000 420}}{31}{16}&0.000 397 5 (18)
\end{tabular}
  \caption{Results for the universal contributions $A_1^{(10)}$.}
  \label{tab:Table3}
\end{table}
In the following, we compare the results obtained in our analysis with
results from previous calculations. The numbers shown are obtained by
numerically integrating over the best available approximation. In case
there are several equivalent approximations the result is obtained by
taking the mean of all values obtained. The errors are then calculated
by taking the difference between the mean and the smallest and largest
values obtained, respectively.

For classes I(a)--I(j) we present our results for the case of purely
electronic vacuum polarization insertions in Tab.~\ref{tab:Table1}. We
compare our results with the values obtained in
Ref.~\cite{Baikov:2012rr}, which relies only on the leading
asymptotics of $\Pi(q^2)$, and results obtained by purely numerical
calculations.

Since the results for classes I(a)-I(c) are obtained by numerical
integrating the exact analytical expression for the one- and two-loop
vacuum polarization they are exact. 

In the case of classes I(d) and I(e) the used three-loop Pad\'es are
highly constrained by a large number of terms in the low- and
high-energy expansion. Thus the error from the spread between different
approximants is negligible. As can be seen the results are in good
agreement with the results obtained in Ref.~\cite{Kinoshita:2005sm} and
the analysis in Ref.~\cite{Baikov:2012rr} can clearly be improved by
including sub-leading contributions.

In case of classes I(f)-I(j) the used four-loop Pad\'es are less
precise but also here we find good agreement within the quoted errors
with the results obtained in Ref.~\cite{Aoyama:2012wk}. In all cases
one finds a significant improvement when comparing to the leading
logarithmic approximation used in Ref.~\cite{Baikov:2012rr}.

For classes I(a)-I(c) and I(e) we can obtain the full result for
$A_2^{(10)}(m_e/m_\mu)$ including muonic contributions. These results
are presented in Tab.~\ref{tab:Table2}. In Tab.~\ref{tab:Table3} we
present our results for the universal corrections and compare with the
results given in Ref.~\cite{Aoyama:2012wj}. In both cases the discussion
as for the purely electronic contributions can essentially be repeated
and also here overall good agreement with results available in the
literature is observed. Nevertheless it should be noted that for single
diagram classes a certain tension remains.

{
As a check of our setup we repeated the analysis of
Ref.~\cite{Baikov:1995ui} and find good agreement 
even though in that reference  a term in the threshold region proportional to
$\log^2(1-z)$ has been omitted (cf. Eq.~(\ref{pi3_thr_coeff_0})). Our
findings at four-loop order are summarized in Tab.~\ref{tab:four-loop}.
\begin{table}%\color{red}
  \centering
  \begin{tabular}{c||c|c|c||}
    &this work&Ref.~\cite{Baikov:1995ui}&Ref.~\cite{Aoyama:2012wk,Kinoshita:2004wi} \\\hline\hline
   I(c) & 1.440741&--&1.440744(16)~\cite{Kinoshita:2004wi}\\\hline
    I(d) &-0.230337&-0.230362(5)&-0.22982(37)~\cite{Aoyama:2012wk} 
  \end{tabular}
  \caption{Comparison of four-loop results for $A_2^{(8)}(m_e/m_\mu)$ obtained in Refs.~\cite{Baikov:1995ui,Aoyama:2012wk} and in this work. The classification of the diagrams corresponds to the notation used in Ref.~\cite{Aoyama:2012wk}.}
  \label{tab:four-loop}
\end{table}
}
% Our new result for diagram class $I(d)$
% the results from
% Refs.~\cite{Baikov:1995ui} and \cite{Aoyama:2012wk}
% \begin{tabular}{ll}
% I(d) B.B. :  &-0.230362(5)\\
% I(d) Kin :   &-0.22982(37)\\
% I(d) B.M.M.: &-0.230337
% \end{tabular}  

% {\color{red}
% The following has to be rephrased....

% As a check of our setup we repeated the analysis of
% Ref.~\cite{Baikov:1995ui} and found complete agreement with their work.

% Internal note: At four loops the numbers are as follows:

% \begin{tabular}{ll}
% I(d) B.B. :  &-0.230362(5)\\
% I(d) Kin :   &-0.22982(37)\\
% I(d) B.M.M.: &-0.230337
% \end{tabular}

% How do we comment on these numbers ?

% For I(c) + I(d) we get 1.21040475\\
% I do not have a number to compare this with, does anybody know the
% correct reference?

% }
\section{Conclusions}
\label{sec:conclusions}
We presented results for a certain set of five-loop contributions to
the anomalous magnetic moment of the muon that stem from corrections
to the vacuum polarization of the photon. We have shown that an
improved treatment of the vacuum polarization, including more than its
asymptotic form, leads to a significantly better agreement with
results obtained by purely numerical methods. It can be clearly seen
that for certain classes of diagrams the asymptotic form of the vacuum
polarization function is not sufficient and power suppressed terms
play an important role and have to be included in the analysis.

\section*{Acknowledgements}
\label{sec:acknowledgements}
We like to thank K.G.~Chetyrkin and J.H.~K\"uhn for initiating the
project, fruitful discussions, and reading of the manuscript.  In
addition, we thank K.G.~Chetyrkin for valuable advice on computing the
vacuum polarization function in high-energy limit.  The work of
P.B. was supported by RFBR grant 11-02-01196. P.M. has been supported
in part by DFG Sonderforschungsbereich Transregio 9,
Computergest\"utzte Theoretische Teilchenphysik, and by the EU Network
{\sf LHCPHENOnet} PITN-GA-2010-264564.

%%%%%%%%%%%%%%%%%%%%%
%  Referenzen
%%%%%%%%%%%%%%%%%%%%%

\bibliography{biblio}{}

\begin{thebibliography}{10}
\expandafter\ifx\csname url\endcsname\relax
  \def\url#1{\texttt{#1}}\fi
\expandafter\ifx\csname urlprefix\endcsname\relax\def\urlprefix{URL }\fi
\expandafter\ifx\csname href\endcsname\relax
  \def\href#1#2{#2} \def\path#1{#1}\fi

\bibitem{Schwinger:1948iu}
J.~S. Schwinger, On quantum electrodynamics and the magnetic moment of the
  electron, Phys.Rev. 73 (1948) 416--417.
\newblock \href {http://dx.doi.org/10.1103/PhysRev.73.416}
  {\path{doi:10.1103/PhysRev.73.416}}.

\bibitem{Petermann:1957hs}
A.~Petermann, {Fourth order magnetic moment of the electron}, Helv.Phys.Acta 30
  (1957) 407--408.

\bibitem{Sommerfield:1957zz}
C.~M. Sommerfield, Magnetic dipole moment of the electron, Phys.Rev. 107 (1957)
  328--329.
\newblock \href {http://dx.doi.org/10.1103/PhysRev.107.328}
  {\path{doi:10.1103/PhysRev.107.328}}.

\bibitem{Kinoshita:1995ym}
T.~Kinoshita, New value of the {$\alpha^3$} electron anomalous magnetic moment,
  Phys.Rev.Lett. 75 (1995) 4728--4731.
\newblock \href {http://dx.doi.org/10.1103/PhysRevLett.75.4728}
  {\path{doi:10.1103/PhysRevLett.75.4728}}.

\bibitem{Laporta:1996mq}
S.~Laporta, E.~Remiddi, The analytical value of the electron (g-2) at order
  {$\alpha^3$} in {QED}, Phys.Lett. B379 (1996) 283--291.
\newblock \href {http://arxiv.org/abs/hep-ph/9602417}
  {\path{arXiv:hep-ph/9602417}}, \href
  {http://dx.doi.org/10.1016/0370-2693(96)00439-X}
  {\path{doi:10.1016/0370-2693(96)00439-X}}.

\bibitem{Melnikov:2000qh}
K.~Melnikov, T.~v. Ritbergen, The three loop relation between the
  {$\overline{\text{MS}}$} and the pole quark masses, Phys.Lett. B482 (2000)
  99--108.
\newblock \href {http://arxiv.org/abs/hep-ph/9912391}
  {\path{arXiv:hep-ph/9912391}}, \href
  {http://dx.doi.org/10.1016/S0370-2693(00)00507-4}
  {\path{doi:10.1016/S0370-2693(00)00507-4}}.

\bibitem{Marquard:2007uj}
P.~Marquard, L.~Mihaila, J.~Piclum, M.~Steinhauser, {Relation between the pole
  and the minimally subtracted mass in dimensional regularization and
  dimensional reduction to three-loop order}, Nucl.Phys. B773 (2007) 1--18.
\newblock \href {http://arxiv.org/abs/hep-ph/0702185}
  {\path{arXiv:hep-ph/0702185}}, \href
  {http://dx.doi.org/10.1016/j.nuclphysb.2007.03.010}
  {\path{doi:10.1016/j.nuclphysb.2007.03.010}}.

\bibitem{Aoyama:2012wk}
T.~Aoyama, M.~Hayakawa, T.~Kinoshita, M.~Nio, Complete tenth-order {QED}
  contribution to the muon g-2, Phys.Rev.Lett. 109 (2012) 111808.
\newblock \href {http://arxiv.org/abs/1205.5370} {\path{arXiv:1205.5370}},
  \href {http://dx.doi.org/10.1103/PhysRevLett.109.111808}
  {\path{doi:10.1103/PhysRevLett.109.111808}}.

\bibitem{Aoyama:2012wj}
T.~Aoyama, M.~Hayakawa, T.~Kinoshita, M.~Nio, Tenth-order {QED} contribution to
  the electron g-2 and an improved value of the fine structure constant,
  Phys.Rev.Lett. 109 (2012) 111807.
\newblock \href {http://arxiv.org/abs/1205.5368} {\path{arXiv:1205.5368}},
  \href {http://dx.doi.org/10.1103/PhysRevLett.109.111807}
  {\path{doi:10.1103/PhysRevLett.109.111807}}.

\bibitem{Lautrup:1974ic}
B.~Lautrup, E.~de~Rafael, {The anomalous magnetic moment of the muon and
  short-distance behaviour of quantum electrodynamics}, Nucl.Phys. B70 (1974)
  317--350.
\newblock \href {http://dx.doi.org/10.1016/0550-3213(74)90481-7}
  {\path{doi:10.1016/0550-3213(74)90481-7}}.

\bibitem{Kinoshita:1990ur}
T.~Kinoshita, H.~Kawai, Y.~Okamoto, {Asymptotic photon propagator in massive
  QED and the muon anomalous magnetic moment}, Phys.Lett. B254 (1991) 235--240.
\newblock \href {http://dx.doi.org/10.1016/0370-2693(91)90427-R}
  {\path{doi:10.1016/0370-2693(91)90427-R}}.

\bibitem{Kawai:1991wd}
H.~Kawai, T.~Kinoshita, Y.~Okamoto, {Asymptotic photon propagator and higher
  order QED Callan-Symanzik beta function}, Phys.Lett. B260 (1991) 193--198.
\newblock \href {http://dx.doi.org/10.1016/0370-2693(91)90990-8}
  {\path{doi:10.1016/0370-2693(91)90990-8}}.

\bibitem{Faustov:1990zs}
R.~Faustov, A.~Kataev, S.~Larin, V.~Starshenko, The analytical contribution of
  the three loop diagrams with two fermion circles to the photon propagator and
  the muon anomalous magnetic moment, Phys.Lett. B254 (1991) 241--246.
\newblock \href {http://dx.doi.org/10.1016/0370-2693(91)90428-S}
  {\path{doi:10.1016/0370-2693(91)90428-S}}.

\bibitem{Kataev:1991cp}
A.~Kataev, {Renormalization group and the five loop QED asymptotic
  contributions to the muon anomaly}, Phys.Lett. B284 (1992) 401--409.
\newblock \href {http://dx.doi.org/10.1016/0370-2693(92)90452-A,
  10.1016/j.physletb.2012.02.057} {\path{doi:10.1016/0370-2693(92)90452-A,
  10.1016/j.physletb.2012.02.057}}.

\bibitem{Broadhurst:1992za}
D.~J. Broadhurst, A.~Kataev, O.~Tarasov, {Analytical on-shell QED results:
  Three loop vacuum polarization, four loop $\beta$ function and the muon
  anomaly}, Phys.Lett. B298 (1993) 445--452.
\newblock \href {http://arxiv.org/abs/hep-ph/9210255}
  {\path{arXiv:hep-ph/9210255}}, \href
  {http://dx.doi.org/10.1016/0370-2693(93)91849-I}
  {\path{doi:10.1016/0370-2693(93)91849-I}}.

\bibitem{Baikov:1995ui}
P.~Baikov, D.~J. Broadhurst, Three loop {QED} vacuum polarization and the four
  loop muon anomalous magnetic moment{ }\href
  {http://arxiv.org/abs/hep-ph/9504398} {\path{arXiv:hep-ph/9504398}}.

\bibitem{Laporta:1993ds}
S.~Laporta, The analytical contribution of some eighth order graphs containing
  vacuum polarization insertions to the muon (g-2) in {QED}, Phys.Lett. B312
  (1993) 495--500.
\newblock \href {http://arxiv.org/abs/hep-ph/9306324}
  {\path{arXiv:hep-ph/9306324}}, \href
  {http://dx.doi.org/10.1016/0370-2693(93)90988-T}
  {\path{doi:10.1016/0370-2693(93)90988-T}}.

\bibitem{Baikov:2012rr}
P.~Baikov, K.~Chetyrkin, J.~K{\"u}hn, C.~Sturm, The relation between the {QED}
  charge renormalized in {$\overline{\text{MS}}$} and on-shell schemes at four
  loops, the qed on-shell {$\beta$}-function at five loops and asymptotic
  contributions to the muon anomaly at five and six loops, Nucl.Phys. B867
  (2013) 182--202.
\newblock \href {http://arxiv.org/abs/1207.2199} {\path{arXiv:1207.2199}},
  \href {http://dx.doi.org/10.1016/j.nuclphysb.2012.09.018}
  {\path{doi:10.1016/j.nuclphysb.2012.09.018}}.

\bibitem{Lee:2013sx}
R.~Lee, P.~Marquard, A.~V. Smirnov, V.~A. Smirnov, M.~Steinhauser, {Four-loop
  corrections with two closed fermion loops to fermion self energies and the
  lepton anomalous magnetic moment}, JHEP 1303 (2013) 162.
\newblock \href {http://arxiv.org/abs/1301.6481} {\path{arXiv:1301.6481}},
  \href {http://dx.doi.org/10.1007/JHEP03(2013)162}
  {\path{doi:10.1007/JHEP03(2013)162}}.

\bibitem{Jegerlehner:2009ry}
F.~Jegerlehner, A.~Nyffeler, The muon g-2, Phys.Rept. 477 (2009) 1--110.
\newblock \href {http://arxiv.org/abs/0902.3360} {\path{arXiv:0902.3360}},
  \href {http://dx.doi.org/10.1016/j.physrep.2009.04.003}
  {\path{doi:10.1016/j.physrep.2009.04.003}}.

\bibitem{Barbieri:1974nc}
R.~Barbieri, E.~Remiddi, {Electron and Muon 1/2(g-2) from Vacuum Polarization
  Insertions}, Nucl.Phys. B90 (1975) 233.
\newblock \href {http://dx.doi.org/10.1016/0550-3213(75)90645-8}
  {\path{doi:10.1016/0550-3213(75)90645-8}}.

\bibitem{Aoyama:2010zp}
T.~Aoyama, M.~Hayakawa, T.~Kinoshita, M.~Nio, Proper eighth-order
  vacuum-polarization function and its contribution to the tenth-order lepton
  g-2, Phys.Rev. D83 (2011) 053003.
\newblock \href {http://arxiv.org/abs/1012.5569} {\path{arXiv:1012.5569}},
  \href {http://dx.doi.org/10.1103/PhysRevD.83.053003}
  {\path{doi:10.1103/PhysRevD.83.053003}}.

\bibitem{Maier:2009fz}
A.~Maier, P.~Maierh{\"o}fer, P.~Marquard, A.~Smirnov, {Low energy moments of
  heavy quark current correlators at four loops}, Nucl.Phys. B824 (2010) 1--18.
\newblock \href {http://arxiv.org/abs/0907.2117} {\path{arXiv:0907.2117}},
  \href {http://dx.doi.org/10.1016/j.nuclphysb.2009.08.011}
  {\path{doi:10.1016/j.nuclphysb.2009.08.011}}.

\bibitem{Nogueira:1993}
P.~Nogueira, Automatic feynman graph generation, J. Comput. Phys. 105 (1993)
  279--289.

\bibitem{Seidensticker:1999bb}
T.~Seidensticker, {Automatic application of successive asymptotic expansions of
  Feynman diagrams }\href {http://arxiv.org/abs/hep-ph/9905298}
  {\path{arXiv:hep-ph/9905298}}.

\bibitem{Harlander:1997zb}
R.~Harlander, T.~Seidensticker, M.~Steinhauser, Complete corrections of {${\cal
  O}(\alpha \alpha_s)$} to the decay of the {$Z$} boson into bottom quarks,
  Phys.Lett. B426 (1998) 125--132.
\newblock \href {http://arxiv.org/abs/hep-ph/9712228}
  {\path{arXiv:hep-ph/9712228}}, \href
  {http://dx.doi.org/10.1016/S0370-2693(98)00220-2}
  {\path{doi:10.1016/S0370-2693(98)00220-2}}.

\bibitem{vermaseren-form}
J.~Vermaseren, New features of \texttt{FORM}.
\newblock \href {http://arxiv.org/abs/math-ph/0010025}
  {\path{arXiv:math-ph/0010025}}.

\bibitem{crusher}
P.~Marquard, D.~Seidel, Crusher, unpublished.

\bibitem{Laporta:2001dd}
S.~Laporta, {High precision calculation of multiloop Feynman integrals by
  difference equations}, Int.J.Mod.Phys. A15 (2000) 5087--5159.
\newblock \href {http://arxiv.org/abs/hep-ph/0102033}
  {\path{arXiv:hep-ph/0102033}}, \href
  {http://dx.doi.org/10.1016/S0217-751X(00)00215-7}
  {\path{doi:10.1016/S0217-751X(00)00215-7}}.

\bibitem{Chetyrkin:2006dh}
K.~Chetyrkin, M.~Faisst, C.~Sturm, M.~Tentyukov, {$\epsilon$}-finite basis of
  master integrals for the integration-by-parts method, Nucl.Phys. B742 (2006)
  208--229.
\newblock \href {http://arxiv.org/abs/hep-ph/0601165}
  {\path{arXiv:hep-ph/0601165}}, \href
  {http://dx.doi.org/10.1016/j.nuclphysb.2006.02.030}
  {\path{doi:10.1016/j.nuclphysb.2006.02.030}}.

\bibitem{Schroder:2005va}
Y.~Schr{\"o}der, A.~Vuorinen, {High-precision epsilon expansions of
  single-mass-scale four-loop vacuum bubbles}, JHEP 0506 (2005) 051.
\newblock \href {http://arxiv.org/abs/hep-ph/0503209}
  {\path{arXiv:hep-ph/0503209}}, \href
  {http://dx.doi.org/10.1088/1126-6708/2005/06/051}
  {\path{doi:10.1088/1126-6708/2005/06/051}}.

\bibitem{Schroder:2005hy}
Y.~Schr{\"o}der, M.~Steinhauser, {Four-loop decoupling relations for the strong
  coupling}, JHEP 0601 (2006) 051.
\newblock \href {http://arxiv.org/abs/hep-ph/0512058}
  {\path{arXiv:hep-ph/0512058}}, \href
  {http://dx.doi.org/10.1088/1126-6708/2006/01/051}
  {\path{doi:10.1088/1126-6708/2006/01/051}}.

\bibitem{Laporta:2002pg}
S.~Laporta, High precision {$\epsilon$} expansions of massive four loop vacuum
  bubbles, Phys.Lett. B549 (2002) 115--122.
\newblock \href {http://arxiv.org/abs/hep-ph/0210336}
  {\path{arXiv:hep-ph/0210336}}, \href
  {http://dx.doi.org/10.1016/S0370-2693(02)02910-6}
  {\path{doi:10.1016/S0370-2693(02)02910-6}}.

\bibitem{Chetyrkin:2004fq}
K.~Chetyrkin, J.~H. K{\"u}hn, P.~Mastrolia, C.~Sturm, Heavy-quark vacuum
  polarization: First two moments of the {${\cal O}(\alpha_s^3 n_f^2)$}
  contribution, Eur.Phys.J. C40 (2005) 361--366.
\newblock \href {http://arxiv.org/abs/hep-ph/0412055}
  {\path{arXiv:hep-ph/0412055}}, \href
  {http://dx.doi.org/10.1140/epjc/s2005-02151-y}
  {\path{doi:10.1140/epjc/s2005-02151-y}}.

\bibitem{Kniehl:2005yc}
B.~A. Kniehl, A.~V. Kotikov, {Calculating four-loop tadpoles with one non-zero
  mass}, Phys.Lett. B638 (2006) 531--537.
\newblock \href {http://arxiv.org/abs/hep-ph/0508238}
  {\path{arXiv:hep-ph/0508238}}, \href
  {http://dx.doi.org/10.1016/j.physletb.2006.04.057}
  {\path{doi:10.1016/j.physletb.2006.04.057}}.

\bibitem{Schroder:2005db}
Y.~Schr{\"o}der, M.~Steinhauser, Four-loop singlet contribution to the
  electroweak {$\rho$} parameter, Phys.Lett. B622 (2005) 124--130.
\newblock \href {http://arxiv.org/abs/hep-ph/0504055}
  {\path{arXiv:hep-ph/0504055}}, \href
  {http://dx.doi.org/10.1016/j.physletb.2005.06.085}
  {\path{doi:10.1016/j.physletb.2005.06.085}}.

\bibitem{Bejdakic:2006vg}
E.~Bejdakic, Y.~Schr{\"o}der, {Hypergeometric representation of a four-loop
  vacuum bubble}, Nucl.Phys.Proc.Suppl. 160 (2006) 155--159.
\newblock \href {http://arxiv.org/abs/hep-ph/0607006}
  {\path{arXiv:hep-ph/0607006}}, \href
  {http://dx.doi.org/10.1016/j.nuclphysbps.2006.09.040}
  {\path{doi:10.1016/j.nuclphysbps.2006.09.040}}.

\bibitem{Kniehl:2006bf}
B.~A. Kniehl, A.~V. Kotikov, {Heavy-quark QCD vacuum polarisation function:
  Analytical results at four loops}, Phys.Lett. B642 (2006) 68--71.
\newblock \href {http://arxiv.org/abs/hep-ph/0607201}
  {\path{arXiv:hep-ph/0607201}}, \href
  {http://dx.doi.org/10.1016/j.physletb.2006.09.008}
  {\path{doi:10.1016/j.physletb.2006.09.008}}.

\bibitem{Kniehl:2006bg}
B.~Kniehl, A.~Kotikov, A.~Onishchenko, O.~Veretin, Strong-coupling constant
  with flavor thresholds at five loops in the modified minimal-subtraction
  scheme, Phys.Rev.Lett. 97 (2006) 042001.
\newblock \href {http://arxiv.org/abs/hep-ph/0607202}
  {\path{arXiv:hep-ph/0607202}}, \href
  {http://dx.doi.org/10.1103/PhysRevLett.97.042001}
  {\path{doi:10.1103/PhysRevLett.97.042001}}.

\bibitem{Baikov:2005nv}
P.~Baikov, {A Practical criterion of irreducibility of multi-loop Feynman
  integrals}, Phys.Lett. B634 (2006) 325--329.
\newblock \href {http://arxiv.org/abs/hep-ph/0507053}
  {\path{arXiv:hep-ph/0507053}}, \href
  {http://dx.doi.org/10.1016/j.physletb.2006.01.052}
  {\path{doi:10.1016/j.physletb.2006.01.052}}.

\bibitem{Baikov:1996rk}
P.~Baikov, {Explicit solutions of the three loop vacuum integral recurrence
  relations}, Phys.Lett. B385 (1996) 404--410.
\newblock \href {http://arxiv.org/abs/hep-ph/9603267}
  {\path{arXiv:hep-ph/9603267}}, \href
  {http://dx.doi.org/10.1016/0370-2693(96)00835-0}
  {\path{doi:10.1016/0370-2693(96)00835-0}}.

\bibitem{Baikov:2010hf}
P.~Baikov, K.~Chetyrkin, Four loop massless propagators: An algebraic
  evaluation of all master integrals, Nucl.Phys. B837 (2010) 186--220.
\newblock \href {http://arxiv.org/abs/1004.1153} {\path{arXiv:1004.1153}},
  \href {http://dx.doi.org/10.1016/j.nuclphysb.2010.05.004}
  {\path{doi:10.1016/j.nuclphysb.2010.05.004}}.

\bibitem{Smirnov:2010hd}
A.~Smirnov, M.~Tentyukov, Four loop massless propagators: a numerical
  evaluation of all master integrals, Nucl.Phys. B837 (2010) 40--49.
\newblock \href {http://arxiv.org/abs/1004.1149} {\path{arXiv:1004.1149}},
  \href {http://dx.doi.org/10.1016/j.nuclphysb.2010.04.020}
  {\path{doi:10.1016/j.nuclphysb.2010.04.020}}.

\bibitem{Hoang:2000yr}
A.~Hoang, M.~Beneke, K.~Melnikov, T.~Nagano, A.~Ota, et~al., {Top - anti-top
  pair production close to threshold: Synopsis of recent NNLO results},
  Eur.Phys.J.direct C2 (2000) 1.
\newblock \href {http://arxiv.org/abs/hep-ph/0001286}
  {\path{arXiv:hep-ph/0001286}}.

\bibitem{Hoang:2008qy}
A.~H. Hoang, V.~Mateu, S.~Mohammad~Zebarjad, {Heavy Quark Vacuum Polarization
  Function at ${\cal O}(\alpha_s^2)$ and ${\cal O}(\alpha_s^3)$}, Nucl. Phys.
  B813 (2009) 349--369.
\newblock \href {http://arxiv.org/abs/0807.4173} {\path{arXiv:0807.4173}},
  \href {http://dx.doi.org/10.1016/j.nuclphysb.2008.12.005}
  {\path{doi:10.1016/j.nuclphysb.2008.12.005}}.

\bibitem{Kiyo:2009gb}
Y.~Kiyo, A.~Maier, P.~Maierh{\"o}fer, P.~Marquard, Reconstruction of heavy
  quark current correlators at {${\cal O}(\alpha_s^3)$}, Nucl.Phys. B823 (2009)
  269--287.
\newblock \href {http://arxiv.org/abs/0907.2120} {\path{arXiv:0907.2120}},
  \href {http://dx.doi.org/10.1016/j.nuclphysb.2009.08.010}
  {\path{doi:10.1016/j.nuclphysb.2009.08.010}}.

\bibitem{Baker:1975}
G.~A. Baker, Essentials of Pad\'e Approximants, Academic Press, Inc., 1975.

\bibitem{Broadhurst:1993mw}
D.~J. Broadhurst, J.~Fleischer, O.~V. Tarasov, {Two loop two point functions
  with masses: Asymptotic expansions and Taylor series, in any dimension}, Z.
  Phys. C60 (1993) 287--302.
\newblock \href {http://arxiv.org/abs/hep-ph/9304303}
  {\path{arXiv:hep-ph/9304303}}, \href {http://dx.doi.org/10.1007/BF01474625}
  {\path{doi:10.1007/BF01474625}}.

\bibitem{Fleischer:1994ef}
J.~Fleischer, O.~V. Tarasov, {Calculation of Feynman diagrams from their small
  momentum expansion}, Z. Phys. C64 (1994) 413--426.
\newblock \href {http://arxiv.org/abs/hep-ph/9403230}
  {\path{arXiv:hep-ph/9403230}}, \href {http://dx.doi.org/10.1007/BF01560102}
  {\path{doi:10.1007/BF01560102}}.

\bibitem{Broadhurst:1994qj}
D.~J. Broadhurst, et~al., {Two loop gluon condensate contributions to heavy
  quark current correlators: Exact results and approximations}, Phys. Lett.
  B329 (1994) 103--110.
\newblock \href {http://arxiv.org/abs/hep-ph/9403274}
  {\path{arXiv:hep-ph/9403274}}, \href
  {http://dx.doi.org/10.1016/0370-2693(94)90524-X}
  {\path{doi:10.1016/0370-2693(94)90524-X}}.

\bibitem{Chetyrkin:1995ii}
K.~G. Chetyrkin, J.~H. {K\"uhn}, M.~Steinhauser, {Heavy Quark Vacuum
  Polarisation to Three Loops}, Phys. Lett. B371 (1996) 93--98.
\newblock \href {http://arxiv.org/abs/hep-ph/9511430}
  {\path{arXiv:hep-ph/9511430}}, \href
  {http://dx.doi.org/10.1016/0370-2693(95)01593-0}
  {\path{doi:10.1016/0370-2693(95)01593-0}}.

\bibitem{Chetyrkin:1996cf}
K.~Chetyrkin, J.~H. K{\"u}hn, M.~Steinhauser, Three loop polarization function
  and {${\cal O}(\alpha_s^2)$} corrections to the production of heavy quarks,
  Nucl.Phys. B482 (1996) 213--240.
\newblock \href {http://arxiv.org/abs/hep-ph/9606230}
  {\path{arXiv:hep-ph/9606230}}, \href
  {http://dx.doi.org/10.1016/S0550-3213(96)00534-2}
  {\path{doi:10.1016/S0550-3213(96)00534-2}}.

\bibitem{Boughezal:2006uu}
R.~Boughezal, M.~Czakon, T.~Schutzmeier, {Four-Loop Tadpoles: Applications in
  QCD}, Nucl.Phys.Proc.Suppl. 160 (2006) 160--164.
\newblock \href {http://arxiv.org/abs/hep-ph/0607141}
  {\path{arXiv:hep-ph/0607141}}, \href
  {http://dx.doi.org/10.1016/j.nuclphysbps.2006.09.041}
  {\path{doi:10.1016/j.nuclphysbps.2006.09.041}}.

\bibitem{Maier:2007yn}
A.~Maier, P.~Maierh{\"o}fer, P.~Marquard, Higher moments of heavy quark
  correlators in the low energy limit at {${\cal O}(\alpha_s^2)$}, Nucl.Phys.
  B797 (2008) 218--242.
\newblock \href {http://arxiv.org/abs/0711.2636} {\path{arXiv:0711.2636}},
  \href {http://dx.doi.org/10.1016/j.nuclphysb.2007.12.035}
  {\path{doi:10.1016/j.nuclphysb.2007.12.035}}.

\bibitem{Maier:2011jd}
A.~Maier, P.~Marquard, Low- and high-energy expansion of heavy-quark
  correlators at next-to-next-to-leading order, Nucl.Phys. B859 (2012) 1--12.
\newblock \href {http://arxiv.org/abs/1110.5581} {\path{arXiv:1110.5581}},
  \href {http://dx.doi.org/10.1016/j.nuclphysb.2012.01.021}
  {\path{doi:10.1016/j.nuclphysb.2012.01.021}}.

\bibitem{baker:1970}
G.~A. Baker~Jr., J.~L. Gammel, The {Pad\'e} Approximant in Theoretical Physics,
  Vol.~71 of Mathematics in Science and Engineering, Academic Press, Inc.,
  1970, Ch.~1, pp. 1--38.

\bibitem{Kinoshita:2005sm}
T.~Kinoshita, M.~Nio, The tenth-order qed contribution to the lepton g-2:
  Evaluation of dominant $\alpha^5$ terms of muon g-2, Phys.Rev. D73 (2006)
  053007.
\newblock \href {http://arxiv.org/abs/hep-ph/0512330}
  {\path{arXiv:hep-ph/0512330}}, \href
  {http://dx.doi.org/10.1103/PhysRevD.73.053007}
  {\path{doi:10.1103/PhysRevD.73.053007}}.

\bibitem{Aoyama:2008hz}
T.~Aoyama, M.~Hayakawa, T.~Kinoshita, M.~Nio, Tenth-order lepton anomalous
  magnetic moment: Second-order vertex containing two vacuum polarization
  subdiagrams, one within the other, Phys.Rev. D78 (2008) 113006.
\newblock \href {http://arxiv.org/abs/0810.5208} {\path{arXiv:0810.5208}},
  \href {http://dx.doi.org/10.1103/PhysRevD.78.113006}
  {\path{doi:10.1103/PhysRevD.78.113006}}.

\bibitem{Aoyama:2008gy}
T.~Aoyama, M.~Hayakawa, T.~Kinoshita, M.~Nio, N.~Watanabe, Eighth-order
  vacuum-polarization function formed by two light-by-light-scattering diagrams
  and its contribution to the tenth-order electron g-2, Phys.Rev. D78 (2008)
  053005.
\newblock \href {http://arxiv.org/abs/0806.3390} {\path{arXiv:0806.3390}},
  \href {http://dx.doi.org/10.1103/PhysRevD.78.053005}
  {\path{doi:10.1103/PhysRevD.78.053005}}.

\bibitem{Kinoshita:2004wi}
T.~Kinoshita, M.~Nio, {Improved alpha**4 term of the muon anomalous magnetic
  moment}, Phys.Rev. D70 (2004) 113001.
\newblock \href {http://arxiv.org/abs/hep-ph/0402206}
  {\path{arXiv:hep-ph/0402206}}, \href
  {http://dx.doi.org/10.1103/PhysRevD.70.113001}
  {\path{doi:10.1103/PhysRevD.70.113001}}.

\end{thebibliography}
\bibliographystyle{elsarticle-num}
\end{document}